\documentclass[11pt,a4paper]{scrartcl}

\usepackage{amssymb}
\usepackage{amsmath}
\usepackage{amsthm}
\usepackage{hyperref}
\usepackage{cite}
\usepackage{graphicx}
\usepackage[disable]{todonotes}
\usepackage{enumitem}
\usepackage{tikz}
\usepackage{booktabs}
\usepackage{multirow}
\usepackage{subfig}
\usetikzlibrary{automata}
\usetikzlibrary{positioning}
\usetikzlibrary{calc}

\newcommand{\VA}[1]{\mathsf{VA}(#1)}
\newcommand{\VAl}[1]{\mathsf{VA}^\emptyWord(#1)}
\newcommand{\VT}[2]{\mathsf{VT}(#1, #2)}
\newcommand{\VTl}[2]{\mathsf{VT}^\emptyWord(#1, #2)}
\newcommand{\NP}{\mathsf{NP}}
\newcommand{\PMon}{\mathcal{C}}
\newcommand{\Parikh}[1]{\Psi(#1)}
\newcommand{\SLMap}[1]{\Phi(#1)}
\newcommand{\TrivMon}{\mathbf{1}}

\newcommand{\monoidsubsetfont}[1]{\mathsf{#1}}
\newcommand{\Rio}{\monoidsubsetfont{R}}
\newcommand{\Lio}{\monoidsubsetfont{L}}
\newcommand{\Uo}{\monoidsubsetfont{H}}
\newcommand{\Eo}{\monoidsubsetfont{J}}
\newcommand{\Xo}{\monoidsubsetfont{X}}
\newcommand{\Ri}[1]{\Rio(#1)}
\newcommand{\Li}[1]{\Lio(#1)}
\newcommand{\U}[1]{\Uo(#1)}
\newcommand{\E}[1]{\Eo(#1)}
\newcommand{\X}[1]{\Xo(#1)}

\newcommand{\RAT}[1]{\mathsf{RAT}(#1)}
\newcommand{\SL}[1]{\mathsf{SL}(#1)}

\newcommand{\LFamily}{\mathcal{F}}

\DeclareMathOperator{\bin}{bin}
\DeclareMathOperator{\dep}{dep}

\newcommand{\F}{\mathcal{F}}
\newcommand{\C}{\mathcal{C}}
\newcommand{\Z}{\mathbb{Z}}
\newcommand{\B}{\mathbb{B}}
\newcommand{\N}{\mathbb{N}}
\newcommand{\M}{\mathbb{M}}
\newcommand{\T}{\mathbb{T}}

\newcommand{\defeq}{=}

\newcommand{\One}{\mathbf{1}}
\newcommand{\step}[1]{\Rightarrow_{#1}}

\newcommand{\congruence}{\equiv}

\newcommand{\emptyWord}{\lambda}
\newcommand{\anEmpty}{a} 
\newcommand{\spelling}{spelling}
\newcommand{\lelim}{strongly $\emptyWord$-independent}

\newcommand{\emphasize}[1]{\paragraph{#1}}

\newenvironment{proofqed}[1]
{
	\begin{proof}
}
{
	\end{proof}
}

\newenvironment{proofqedtitle}[1]
{
	\begin{proof}[#1]
}
{
	\end{proof}
}

\newenvironment{proofqednotitle}
{
	\begin{proof}
}
{
	\end{proof}
}

\newcommand{\loopedpath}[1]{
\begin{tikzpicture}[every circle/.style={}, scale=#1]
\fill (0,0) circle (2pt) node (a) {}    (1,0) circle (2pt) node (b)  {}   (2,0) circle (2pt) node (c) {}   (3,0) circle (2pt) node (d) {};
\draw (a.center) -- (b.center) -- (c.center) -- (d.center);
\draw (b.center) ++(90:3pt) circle (3pt);
\draw (c.center) ++(90:3pt) circle (3pt);
\end{tikzpicture}
}

\newcommand{\blindcounters}[1]{
\begin{tikzpicture}[every circle/.style={}, scale=#1]
\fill (0:0) circle (2pt) node (aa) {}    +(60:1) circle (2pt) node (ab) {}      +(0:1) circle (2pt) node (ac) {} ;
\draw (aa.center) -- (ab.center) -- (ac.center) -- (aa.center);
\draw (aa.center) ++(-150:3pt) circle (3pt);
\draw (ab.center) ++(90:3pt) circle (3pt);
\draw (ac.center) ++(-30:3pt) circle (3pt);
\end{tikzpicture}
}

\newcommand{\pushdownstorage}[1]{
\begin{tikzpicture}[every circle/.style={}, scale=#1]
\fill (0:0) circle (2pt) node (aa) {}    +(60:1) circle (2pt) node (ab) {}      +(0:1) circle (2pt) node (ac) {} ;
\end{tikzpicture}
}

\newcommand{\partiallyblindcounters}[1]{
\begin{tikzpicture}[every circle/.style={}, scale=#1]
\fill (0:0) circle (2pt) node (aa) {}    +(60:1) circle (2pt) node (ab) {}      +(0:1) circle (2pt) node (ac) {} ;
\draw (aa.center) -- (ab.center) -- (ac.center) -- (aa.center);
\end{tikzpicture}
}

\newcommand{\pushdowncounters}[1]{
\begin{tikzpicture}[every circle/.style={}, scale=#1]
\fill (0,0) circle (2pt) node (aa) {}    +(1,0) circle (2pt) node (ab) {}      +(1,1) circle (2pt) node (ac) {}    +(0,1) circle (2pt) node (ad) {};
\draw (aa.center) -- (ab.center) -- (ac.center) -- (ad.center);
\draw (ad.center) -- (ab.center);
\draw (aa.center) -- (ac.center);
\draw (ab.center) ++(-45:3pt) circle (3pt);
\draw (ac.center) ++(45:3pt) circle (3pt);
\end{tikzpicture}
}

\newcommand{\depgraphzrightx}[1]{
\begin{tikzpicture}[every circle/.style={}, scale=#1]
\path node (x) at (0,0) {$x$}    
      node (y) at +(30:1)  {$y$} 
      node (ys) at +(-30:1)  {$y'$}
      node (z) at +(2,0) {$z$};
\draw[->,dashed] (z) to (x);
\draw[->] (y) to  [out=0,  in=90] (z);
\draw[->] (y) to (ys);
\end{tikzpicture}
}

\newcommand{\depgraphzrightyp}[1]{
\begin{tikzpicture}[every circle/.style={}, scale=#1]
\path node (x) at (0,0) {$x$}    
      node (y) at +(30:1)  {$y$} 
      node (ys) at +(-30:1)  {$y'$}
      node (z) at +(2,0) {$z$};
\draw[->] (x) to (z);
\draw[->] (y) to (ys);
\draw[->] (y) to  [out=0,  in=90] (z);
\draw[->,dashed] (z) to [out=-90, in=0] (ys);
\end{tikzpicture}
}

\newcommand{\depgraphzleftx}[1]{
\begin{tikzpicture}[every circle/.style={}, scale=#1]
\path node (x) at (0,0) {$x$}    
      node (y) at +(30:1) {$y$} 
      node (ys) at +(-30:1) {$y'$}
      node (z) at +(-1,0) {$z$};
\draw[->,dashed] (x) to (z);
\draw[->] (y) to (ys);
\draw[->] (z) to [out=90, in=-180] (y);
\end{tikzpicture}
}

\newcommand{\depgraphzleftyp}[1]{
\begin{tikzpicture}[every circle/.style={}, scale=#1]
\path node (x) at (0,0) {$x$}    
      node (y) at +(30:1) {$y$} 
      node (ys) at +(-30:1) {$y'$}
      node (z) at +(-1,0) {$z$};
\draw[->] (z) to (x);
\draw[->] (y) to (ys);
\draw[->] (z) to [out=90, in=-180] (y);
\draw[->,dashed] (ys) to [out=180, in=-90] (z);
\end{tikzpicture}
}

\newcommand{\depgraphxydep}[1]{
\begin{tikzpicture}[every circle/.style={}, scale=#1]
\path node (x) at (0,0) {$x$}    
      node (y) at +(30:1) {$y$} 
      node (ys) at +(-30:1) {$y'$};
\draw[->,dashed] (x) to (y);
\draw[->] (y) to (ys);
\draw[->,dashed] (x) to (ys);
\end{tikzpicture}
}

\newtheorem{theorem}{Theorem}[section]
\newtheorem{lemma}[theorem]{Lemma}
\newtheorem{corollary}[theorem]{Corollary}
\theoremstyle{definition}
\newtheorem{definition}[theorem]{Definition}

\begin{document}
\title{Silent Transitions in Automata with Storage}

\author{Georg Zetzsche}
\titlehead{Fachbereich Informatik, Technische Universit\"{a}t Kaiserslautern, \\
Postfach 3049, 67653 Kaiserslautern, Germany \\
\texttt{zetzsche@cs.uni-kl.de}}

\maketitle

\begin{abstract}
We consider the computational power of silent transitions in one-way automata
with storage.  Specifically, we ask which storage mechanisms admit a
transformation of a given automaton into one that accepts the same language
and reads at least one input symbol in each step.

We study this question using the model of valence automata.  Here, a finite
automaton is equipped with a storage mechanism that is given by a monoid.

This work presents generalizations of known results on silent transitions.
For two classes of monoids, it provides characterizations of those monoids that
allow the removal of $\emptyWord$-transitions.  Both classes are defined by
graph products of copies of the bicyclic monoid and the group of integers.  The
first class contains pushdown storages as well as the blind counters while the
second class contains the blind and the partially blind counters.
\end{abstract}

\listoftodos

\section{Introduction}
We consider the problem of removing silent transitions from one-way
automata with various kinds of storage.  Specifically, we ask for which kinds
of storage the real-time and the general version have equal computational power.

This is an interesting problem for two reasons. First, it has consequences for
the time and space complexity of the membership problem for these automata. For
automata with silent transitions, it is not even clear whether the
membership problem is decidable.  If, however,  an automaton has no
silent transitions, we only have to consider paths that are at most as long as
the word at hand.  In particular, if we can decide whether a sequence of
storage operations is valid using linear space, we can also solve the
membership problem (nondeterministically) with a linear space bound. Similarly, if we can decide validity of
such a sequence in polynomial time, we can solve the membership problem in (nondeterministic) polynomial time. 

Second, we can interpret the problem as a question on resource consumption 
of restricted machine models: we ask for which storage mechanisms we can 
process every input word by executing only a bounded number of operations per symbol.

\todo{Add reference for Higher-Order Pushdown}
There is a wide variety of machine models that consist of a finite state
control with a one-way input and some mechanism to store data, for example 
(higher order) pushdown automata, various kinds of counter automata \cite{Greibach1978}, or
off-line Turing machines that can only move right on the input tape.

For some of these models, it is known whether $\emptyWord$-transitions can be
eliminated.  For example, the Greibach normal form allows their removal from
\emph{pushdown automata} \cite{Greibach1965}. Furthermore,
for \emph{blind counter automata} (i.e., the counters can go below zero and a
zero-test is only performed in the end), Greibach also has also shown that
$\emptyWord$-transitions can be avoided \cite{Greibach1978}. However, for
\emph{partially blind counter automata} (i.e., the counters cannot go below
zero and are only zero-tested in the end) or, equivalently, Petri nets, there
are languages for which $\emptyWord$-transitions are indeed necessary
\cite{Greibach1978,Jantzen1979a,Jantzen1979b}. 

The aim of this work is to generalize these results and obtain insights into
how the properties of the storage mechanism influence the computational power
of the real-time variant. 

In order to study the expressive power of real-time computations in greater
generality, we use the model of \emph{valence automata}.  For our purposes, a
storage mechanism consists of a (possibly infinite) set of states and partial
transformations operating on them. Such a mechanism often works in a way such
that a computation is considered valid if the composition of the applied
transformations is the identity. For example, in a pushdown storage, the
operations \emph{push} and \emph{pop} (for each participating stack symbol) and
compositions thereof are partial transformations on the set of words over some
alphabet. In this case, a computation is valid if, in the end, the stack is
brought back to the initial state, i.e., the identity transformation has been
applied. Furthermore, in a partially blind counter automaton, a computation is
valid if it leaves the counters with value zero, i.e., the composition of the
applied operations \emph{increase} and \emph{decrease} is the identity.
Therefore, the set of all compositions of the partial transformations forms a
monoid such that in many cases, a computation is valid if the composition of
the transformations is the identity map.

A valence automaton is a finite automaton in which each edge carries, in
addition to an input word, an element of a monoid.  A word is then accepted if
there is a computation that spells the word and for which the product of the
monoid elements is the identity.  Valence automata have been studied throughout
the last decades
\cite{IbarraSahniKim1976,Gilman1996,ItoMartinVideMitrana2001,MitranaStiebe2001,ElstonOstheimer2004, ElderKambitesOstheimer2008,Kambites2009,RenderKambites2009,Render2010}.

The contribution of this work is threefold. On the one hand, we introduce a
class of monoids that accommodates, among others, all storage mechanisms for
which we mentioned previous results on silent transitions.  The monoids in this
class are graph products of copies of the bicyclic monoid and the integers.  On
the other hand, we present two generalizations of those established facts.  Our
first main result is a characterization of those monoids in a certain subclass
for which $\emptyWord$-transitions can be eliminated. This subclass contains,
among others, both the monoids corresponding to pushdown storages as well as
those corresponding to blind multicounter storages. Thus, we obtain a
generalization and unification of two of the three $\emptyWord$-removal results
above.  For those storage mechanisms in this subclass for which we can remove
$\emptyWord$-transitions, there is a simple intuitive description.

The second main result is a characterization of the previous kind for the class
of those storage mechanisms that consist of a number of blind counters and a
number of partially blind counters. Specifically, we show that we can remove
$\emptyWord$-transitions if and only if there is at most one partially blind
counter. Again, this generalizes and unifies two of the three results above.

The rest of the paper is organized as follows. In Section \ref{sec:notions}, we
will fix notation and define some basic concepts. In Section \ref{sec:results}, we state
the main results, describe how they relate to what is known, and explain key
ideas.  Sections \ref{sec:semilinear}, \ref{sec:membership}, and
\ref{sec:rational} contain auxiliary results needed in Section \ref{sec:silent},
which presents the proofs of the main results.

\section{Basic Notions}
\label{sec:notions}
We assume that the reader has some basic knowledge on formal languages and monoids.
In this section, we will fix some notation and introduce basic concepts.

A \emph{monoid} is a set $M$ together with an associative operation and a
neutral element.  Unless defined otherwise, we will denote the neutral element
of a monoid by $1$ and its operation by juxtaposition.  That is, for a monoid
$M$ and $a,b\in M$, $ab\in M$ is their product.  
For $a,b\in M$, we write
$a\sqsubseteq b$ if there is a $c\in M$ such that $b=ac$.   By
$\One$, we denote the trivial monoid that consists of just one element.

We call a monoid \emph{commutative} if $ab=ba$ for any $a,b\in M$.
A subset $N\subseteq M$ is said to be a \emph{submonoid of $M$} if $1\in
N$ and $a,b\in N$ implies $ab\in N$. For a subset $N\subseteq M$, let
$\langle N\rangle$ be the intersection of all submonoids $N'$ of $M$ that
contain $N$. That is, $\langle N\rangle$ is the smallest submonoid of $M$
that contains $N$.  $\langle N\rangle$ is also called the \emph{submonoid
generated by $N$}. In each monoid $M$, we have the following submonoids:
\begin{eqnarray*}
\U{M}&\defeq& \{a\in M \mid \exists b\in M: ab=ba=1 \}, \\
\Ri{M}&\defeq& \{a\in M \mid \exists b\in M: ab=1 \}, \\
\Li{M}&\defeq& \{a\in M \mid \exists b\in M: ba=1 \}.
\end{eqnarray*}
When using a monoid $M$ as part of a control mechanism, the subset 
\[ \E{M}\defeq \{a\in M \mid \exists b,c\in M: bac=1\} \] 
will play an important role.
By $M^n$, we denote the $n$-fold direct product
of $M$, i.e.  $M^n=M\times\cdots\times M$ with $n$ factors.

Let $S\subseteq M$ be a subset. If there is no danger of confusion with the
$n$-fold direct product, we write $S^n$ for the set of all elements of $M$
that can be written as a product of $n$ factors from $S$.

Let $\Sigma$ be a fixed countable set of abstract symbols, the finite subsets
of which are called \emph{alphabets}.  For an alphabet $X$, we will write $X^*$
for the set of words over $X$.  The empty word is denoted by $\emptyWord\in
X^*$.  Together with the concatenation as its operation, $X^*$ is a monoid.  We
will regard every $x\in X$ as an element of $X^*$, namely the word consisting
of only one occurrence of $x$.  For a symbol $x\in X$ and a word $w\in X^*$,
let $|w|_x$ be the number of occurrences of $x$ in $w$. For a subset
$Y\subseteq X$, let $|w|_Y\defeq\sum_{x\in Y}|w|_x$.  By $|w|$, we will refer
to the length of $w$.   Given an alphabet $X$ and a monoid $M$, subsets of
$X^*$ and $X^*\times M$ are called \emph{languages} and \emph{transductions},
respectively. A \emph{family} is a set of languages that is closed under
isomorphism and contains at least one non-trivial member. For a subset
$Y\subseteq X$, we define the homomorphism $\pi_Y:X^*\to Y^*$ by $\pi_Y(y)=y$
for $y\in Y$ and $\pi_Y(x)=\emptyWord$ for $x\in X\setminus Y$.

Given an alphabet $X$, we write $X^\oplus$ for the set of maps $\alpha:X\to\N$.
Elements of $X^\oplus$ are called \emph{multisets}.  By way of pointwise
addition, written $\alpha+\beta$, $X^\oplus$ is a commutative monoid (also called 
the \emph{free commutative monoid} over $X$).  We write
$0$ for the empty multiset, i.e. the one that maps every $x\in X$ to $0\in\N$.
For $\alpha\in X^\oplus$, let $|\alpha|=\sum_{x\in X} \alpha(x)$.  The
\emph{Parikh mapping} is the mapping $\Psi:\Sigma^*\to\Sigma^\oplus$ defined by
$\Psi(w)(x)\defeq |w|_x$ for all $w\in\Sigma^*$ and $x\in\Sigma$. 

Let $A$ be a (not necessarily finite) set of symbols and $R\subseteq A^*\times
A^*$.  The pair $(A,R)$ is called a \emph{(monoid) presentation}. The smallest
congruence of $A^*$ containing $R$ is denoted by $\congruence_R$ and we will write
$[w]_R$ for the congruence class of $w\in A^*$.  The \emph{monoid presented by
$(A,R)$} is defined as $A^*/\mathord{\congruence_R}$. Note that since we did not impose a
finiteness restriction on $A$, every monoid has a presentation. Furthermore,
for monoids $M_1$, $M_2$ we can find presentations $(A_1,R_1)$ and $(A_2,R_2)$
such that $A_1\cap A_2=\emptyset$.  We define the \emph{free product} $M_1*M_2$
to be presented by $(A_1\cup A_2, R_1\cup R_2)$. Note that $M_1*M_2$ is
well-defined up to isomorphism. By way of the injective morphisms
$[w]_{R_i}\mapsto [w]_{R_1\cup R_2}$, $w\in A_i^*$ for $i=1,2$, we will
regard $M_1$ and $M_2$ as subsets of $M_1*M_2$.
In analogy to the $n$-fold direct product, we
write $M^{(n)}$ for the $n$-fold free product of $M$.

\emphasize{Rational Sets}
Let $M$ be a monoid. An \emph{automaton over $M$} is a tuple $A=(Q,M,E,q_0,F)$, in which 
$Q$ is a finite set of \emph{states}, $E$ is a finite
subset of $Q\times M\times Q$ called the set of \emph{edges}, $q_0\in Q$ is the \emph{initial
state}, and $F\subseteq Q$ is the set of \emph{final states}. The \emph{step
relation} $\step{A}$ of $A$ is a binary relation on $Q\times M$, for
which $(p,a) \step{A} (q,b)$ iff there is an edge $(p,c,q)$ such that
$b=ac$. The set generated by $A$ is then
\[S(A)\defeq\{a\in M \mid \exists q\in F: (q_0,1)\step{A}^* (q,a) \}. \]

A set $R\subseteq M$ is called \emph{rational} if it can be written as $R=S(A)$
for some automaton $A$ over $M$. The set of rational subsets of $M$ is denoted by $\RAT{M}$.
Given two subsets $S,T\subseteq M$, we define $ST=\{st \mid s\in S, t\in T\}$. 
Since $\{1\}\in\RAT{M}$ and $ST\in\RAT{M}$ whenever $S,T\in\RAT{M}$, this
operation makes $\RAT{M}$ a monoid itself.

Let $C$ be a commutative monoid for which we write the composition additively.
For $n\in\N$ and $c\in C$, we use $nc$ to denote $c+\cdots+c$ ($n$ summands).
A subset $S\subseteq C$ is \emph{linear} if there are elements $s_0,\ldots,s_n$
such that $S=\{s_0+\sum_{i=1}^n a_is_i \mid a_i\in\N,~ 1\le i\le n\}$.  A set $S\subseteq C$
is called \emph{semilinear} if it is a finite union of linear sets.  By
$\SL{C}$, we denote the set of semilinear subsets of $C$. It is well-known that
$\RAT{C}=\SL{C}$ for commutative $C$ (we will, however, sometimes still use
$\SL{C}$ to make explicit that the sets at hand are semilinear). Moreover,
$\SL{C}$ is a commutative monoid by way of the product $(S,T)\mapsto S+T=\{s+t
\mid s\in S, t\in T\}$. It is well-known that the class of semilinear subsets of a free 
commutative monoid is closed under Boolean operations (see
\cite{GinsburgSpanier1964,GinsburgSpanier1966}).

In slight
abuse of terminology, we will sometimes call a language $L$ semilinear if the
set $\Parikh{L}$ is semilinear. If there is no danger of confusion, we will write
$S^\oplus$ instead of $\langle S\rangle$ if $S$ is a subset of a commutative monoid $C$.
Note that, if $X$ is regarded as a subset of the set of multisets over $X$, the two meanings
of $X^\oplus$ coincide.

\emphasize{Valence Automata}
A \emph{valence automaton over $M$} is an automaton $A$ over $X^*\times M$,
where $X$ is an alphabet. An edge $(p,w,m,q)$ in $A$ is called \anEmpty{}
\emph{$\emptyWord$-transition} if $w=\emptyWord$. $A$ is called
\emph{$\emptyWord$-free} if it has no $\emptyWord$-transitions.  The
\emph{language accepted by $A$} is defined as
\[ L(A)\defeq \{w\in X^* \mid (w,1)\in S(A)\}. \]
The class of languages accepted by valence automata and $\emptyWord$-free valence
automata over $M$ is denoted by $\VAl{M}$ and $\VA{M}$, respectively.

A \emph{finite automaton} is a valence automaton over the trivial monoid $\One$.
For a finite automaton $A=(Q, X^*\times\One,E,q_0,F)$, we also write
$A=(Q,X,E,q_0,F)$.
Languages accepted by finite automata are called \emph{regular languages}.
The finite automaton $A$ is \emph{\spelling{}}, if $E\subseteq Q\times X\times Q$, i.e. every edges carries
exactly one letter.
Let $M$ and $C$ be monoids.
A \emph{valence transducer over $M$ with output in $C$} is an automaton $A$ over
$X^*\times M\times C$, where $X$ is an alphabet. The \emph{transduction
performed by $A$} is
\[T(A)\defeq \{(x,c)\in X^*\times C \mid (x,1,c)\in S(A) \}.\]
A valence transducer is called \emph{$\emptyWord$-free} if it is $\emptyWord$-free as a
valence automaton.  We denote the class of transductions performed by
($\emptyWord$-free) valence transducers over $M$ with output in $C$ by
$\VTl{M}{C}$ ($\VT{M}{C}$).

\emphasize{Graphs}
A \emph{graph} is a pair $\Gamma=(V,E)$ where $V$ is a finite set and
$E\subseteq \{S\subseteq V \mid |S|\le 2\}$.  The elements of $V$ are called
\emph{vertices} and those of $E$ are called \emph{edges}.  If $\{v\}\in E$ for some $v\in V$,
then $v$ is called a \emph{looped} vertex and otherwise it is \emph{unlooped}.
A \emph{subgraph} of $\Gamma$ is a graph $(V',E')$ with $V'\subseteq V$ and  $E'\subseteq E$. 
Such a subgraph is called
\emph{induced (by $V'$)} if $E'=\{ S\in E \mid S\subseteq V'\}$, i.e. $E'$
contains all edges from $E$ incident to vertices in $V'$.  By
$\Gamma\setminus\{v\}$, for $v\in V$, we denote the subgraph of $\Gamma$
induced by $V\setminus \{v\}$.  Given a graph $\Gamma=(V,E)$, its
\emph{underlying loop-free graph} is $\Gamma'=(V,E')$ with $E'=E\cap
\{S\subseteq V \mid |S|=2 \}$.  For a vertex $v\in V$, the elements of
$N(v)=\{w\in V \mid \{v,w\}\in E\}$ are called \emph{neighbors} of $v$.  A
\emph{looped clique} is a graph in which $E=\{S\subseteq V \mid |S|\le 2\}$.
Moreover, a \emph{clique} is a loop-free graph in which any two distinct
vertices are adjacent. Finally, an \emph{anti-clique} is a graph with
$E=\emptyset$.

A presentation $(A,R)$ in which $A$ is a finite alphabet is a \emph{Thue system}.
To each graph $\Gamma=(V,E)$, we associate the Thue system $T_\Gamma=(X_\Gamma, R_\Gamma)$ over the alphabet 
$X_\Gamma=\{a_v, \bar{a}_v \mid v\in V\}$. $R_\Gamma$ is defined as
\[ R_\Gamma = \{(a_v\bar{a}_v, \emptyWord) \mid v\in V \} \cup \{ (xy,yx) \mid x\in \{a_v, \bar{a}_v\}, ~y\in \{a_w,\bar{a}_w\}, ~ \{v,w\}\in E\}. \]
In particular, we have $(a_v\bar{a}_v, \bar{a}_va_v)\in R_\Gamma$ whenever $\{v\}\in E$.
To simplify notation, the congruence $\congruence_{T_\Gamma}$ is then also denoted by $\congruence_\Gamma$.
In order to describe the monoids we use to model storage mechanisms, we define monoids using graphs.
To each graph $\Gamma$, we associate the monoid
\[ \M\Gamma ~~=~~ X^*_\Gamma/\mathord{\congruence_\Gamma}. \]

If $\Gamma$ consists of one vertex and has no edges, $\M\Gamma$ is also denoted
as $\B$ and we will refer to it as the \emph{bicyclic monoid}. The generators
$a_v$ and $\bar{a}_v$ are then also written $a$ and $\bar{a}$, respectively.

\section{Results}
\label{sec:results}

\begin{table}
\begin{center}
\newcommand{\drawfactor}{1}
\newcommand{\seperator}{}
\begin{tabular}{ccl} \toprule
Graph $\Gamma$ & Monoid $\M\Gamma$ & Storage mechanism \\\midrule
\multirow{3}{*}{\pushdownstorage{\drawfactor}}        &            &  \\
                                            & $\B^{(3)}$ & Pushdown (with three symbols) \\
                                            &            &                  \\ \seperator
\multirow{3}{*}{\partiallyblindcounters{\drawfactor}} &            & \\
                                            & $\B^{3}$   & Three partially blind counters \\
                                            &            & \\ \seperator
\multirow{3}{*}{\blindcounters{\drawfactor}}          &            & \\
                                            & $\Z^3$     & Three blind counters \\
				            &            &                \\ \seperator
\multirow{3}{*}{\pushdowncounters{\drawfactor}}          &            & \\
                                            & $\B^{(2)}\times \Z^2$     & Pushdown (with two symbols) and two blind counters \\
				            &            &                \\ \bottomrule
\end{tabular}
\end{center}
\caption{Examples of storage mechanisms}
\label{tab:graphexamples}
\end{table}

\emphasize{Storage mechanisms as monoids}
First of all, we will see how pushdown storages and (partially) blind
counters can be regarded as monoids of the form $\M\Gamma$. See Table \ref{tab:graphexamples} for a set of examples. It is not hard to see that in
the bicyclic monoid $\B$, a word over the generators $a$ and $\bar{a}$ is the identity if
and only if in every prefix of the word, there are at least as many $a$'s as there are $\bar{a}$'s
and in the whole word, there are as many $a$'s as there are $\bar{a}$'s. Thus, a valence 
automaton over $\B$ is an automaton with one counter that cannot go below zero and is zero
in the end. Here, the increment operation corresponds to $a$ and the decrement corresponds to
$\bar{a}$. 

Observe that building the direct product means that both storage mechanisms
(described by the factors) are available and can be used simultaneously. Thus,
valence automata over $\B^n$ are automata with $n$ partially blind counters.
Therefore, if $\Gamma$ is a clique, then $\M\Gamma\cong\B^n$ corresponds to a
partially blind multicounter storage.

Furthermore, the free product of a monoid $M$ with $\B$ yields what can be seen
as a stack of elements of $M$: a valence automaton over $M*\B$ can store a
sequence of elements of $M$ (separated by $a$) such that it can only remove the
topmost element if it is the identity element. The available operations are
those available for $M$ (which then operate on the topmost entry) and in
addition \emph{push} (represented by $a$) and \emph{pop} (represented by
$\bar{a}$). Thus, $\B*\B$ corresponds to a stack over two symbols. In
particular, if $\Gamma$ is an anti-clique (with at least two vertices), then
$\M\Gamma\cong\B^{(n)}$ represents a pushdown storage.

Finally, valence automata over $\Z^n$ (regarded as a monoid by way of addition)
correspond to automata with $n$ blind counters. Hence, if $\Gamma$ is a looped
clique, then $\M\Gamma\cong\Z^n$ corresponds to a blind multicounter storage.

\emphasize{Main results}
Our class of monoids that generalizes pushdown and blind multicounter storages
is the class of $\M\Gamma$ where in $\Gamma$, any two looped vertices are
adjacent and any two unlooped vertices are not adjacent. Our first main result
is the following.
\begin{theorem} \label{theorem:firstmain}
Let $\Gamma$ be a graph such that
\begin{itemize}
\item between any two looped vertices, there is an edge, and
\item between any two unlooped vertices, there is no edge.
\end{itemize}
Then the following assertions are equivalent:
\begin{enumerate}[label=(\arabic*)]
\item\label{item:firstmain:equal} $\VA{\M\Gamma}=\VAl{\M\Gamma}$.
\item\label{item:firstmain:cs}  Every language in $\VAl{\M\Gamma}$ is context-sensitive.
\item\label{item:firstmain:np} The membership problem of each language in $\VAl{\M\Gamma}$ is in $\NP$.
\item\label{item:firstmain:decidable} Every language in $\VAl{\M\Gamma}$ is decidable.
\item\label{item:firstmain:nopath} $\Gamma$ does not contain \loopedpath{1} as an induced subgraph.
\end{enumerate}
\end{theorem}
Note that this generalizes the facts that in pushdown automata and in blind
counter automata, $\emptyWord$-transitions can be avoided.

It turns out that the storages that satisfy the equivalent conditions of
Theorem \ref{theorem:firstmain} (and the hypothesis), are exactly those in the
following class.
\begin{definition}
Let $\PMon$ be the smallest class of monoids such that $\TrivMon\in\C$ and
whenever $M\in\C$, we also have $M\times \Z\in \C$ and $M*\B\in \C$.
\end{definition}
Thus, $\C$ contains those storage types obtained by successively \emph{adding blind counters} and
\emph{building a stack of elements}. For example, we could have a stack each of whose entries
contains $n$ blind counters. Or we could have an ordinary pushdown and a number of blind counters.
Or a stack of elements, each of which is a pushdown storage and a blind counter, etc.

Our second main result concerns storages consisting of a number of blind
counters and a number of partially blind counters.
\begin{theorem}\label{theorem:secondmain}
Let $\Gamma$ be a graph such that between any two distinct vertices, there is
an edge.  Then 
\[ \VA{\M\Gamma}=\VAl{\M\Gamma}~~\text{if and only if}~~ r\le 1, \]
where $r$ is the number of unlooped vertices in $\Gamma$.
\end{theorem}
In other words, when you have $r$ partially blind counters and $s$ blind
counters, $\emptyWord$-transitions can be eliminated if and only if $r\le 1$.
Note that this generalizes Greibach's result that in partially blind multicounter automata,
$\emptyWord$-transitions are indispensable.

\emphasize{Key technical ingredients}
As a first step, we show that  for $M\in\C$, all languages in $\VAl{M}$ are
semilinear.  This is needed in various situations throughout the proof. We
prove this using an old result by van Leeuwen~\cite{vanLeeuwen1974}, which says that languages that
are algebraic over a class of semilinear languages are semilinear themselves.
Thereby, the corresponding lemma \ref{lemma:freeproductalg} slightly generalizes one of the central
components in a decidability result by Lohrey and Steinberg on the rational subset
membership problem for graph groups \cite{LohreySteinberg2008} and provides a simpler proof
(relying, however, on van Leeuwen's result).

Second, we use an undecidability result by Lohrey and Steinberg
\cite{LohreySteinberg2008} concerning the rational subset membership problem
for certain graph groups. We deduce that for monoids $M$ outside of $\C$ (and
satisfying the hypothesis of Theorem \ref{theorem:firstmain}), $\VAl{M}$
contains undecidable languages.

Third, in order to prove our claim by induction on the construction of
$M\in\C$, we use a significantly stronger induction hypothesis: we show that it
is not only possible to remove $\emptyWord$-transitions from valence automata,
but also from valence transducers with output in a commutative monoid. 
Here, however, the constructed valence transducer is allowed to output
a semilinear set in each step. Monoids that admit such a transformation will be called \emph{\lelim}.

Fourth, we develop a normal form result for rational subsets of monoids in $\C$
(see section \ref{sec:rational}). Such normal form results have been available
for monoids described by monadic rewriting systems (see, for example,
\cite{BookOtto1993}), which was applied by Render and Kambites to monoids
representing pushdown storages \cite{RenderKambites2009}.  Under different
terms, this normal form trick has been used by Bouajjani, Esparza, and Maler
\cite{BouajjaniEsparzaMaler1997} and by Caucal \cite{Caucal2003} to describe
rational sets of pushdown operations.  However, since the monoids in $\C$ allow
commutation of certain non-trivial elements, a more general technique was necessary
here.  In the case of monadic rewriting systems, one transforms a finite
automaton according to rewriting rules by gluing in new edges. Here, we glue in
automata accepting sets that are semilinear by earlier steps in the proof. See
Lemma \ref{lemma:ratnormal} for details.

Fifth, we have three techniques to eliminate $\emptyWord$-transitions from
valence transducers while retaining the output in a commutative monoid. Here,
we need one technique to show that if 
$M$ is \lelim, then $M\times\Z$ is as well. This technique
again uses the semilinearity of certain sets and a result that provides small
preimages for morphisms from free commutative monoids to the integers.

The second technique is to show that $\B$ is \lelim. Here, we use a construction that
allows the postponement of increment operations and the early execution of
decrement operations.  This is then used to show that one can restrict oneself
to computations in which a sequence of increments, followed by a sequence of
decrements, will in the end change the counter only by a bounded amount.

The third technique is to show that if 
$M$ is \lelim, where $M$ is nontrivial, then 
$M*\B$ is as well. Here, the storage consists of a stack of elements of $M$. The
construction works by encoding rational sets over $M*\B$ as elements on the
stack. We have to use the semilinearity results again in order to be able to
compute the set of all possible outputs when elements from two given rational
sets cancel each other out (in the sense that push operations are followed by
pop operations).

\section{Semilinear Languages}
\label{sec:semilinear}

This section contains semilinearity results that will be needed in later sections.
The first lemma guarantees small preimages of morphisms from multisets to the integers.
This will be used to bound the number of necessary operations on a blind counter in order
to obtain a certain counter value.
\begin{lemma}\label{lemma:preimages}
Let $\varphi:X^\oplus \to \Z$ be a morphism. Then for any $n\in\Z$, the set
$\varphi^{-1}(n)$ is semilinear. In particular, $\ker\varphi$ is finitely generated.
Furthermore, there is a constant $k\in\N$ such that for any $\mu\in X^\oplus$, 
there is a $\nu\sqsubseteq\mu$ with $\mu\in\nu+\ker\varphi$ and $|\nu|\le k\cdot |\varphi(\mu)|$.
\end{lemma}
\begin{proofqed}{lemma:preimages}
It is clearly possible to construct a context-free grammar for the language
$L=\{w\in\{a,b\}^* \mid |w|_a-|w|_b=n \}$. Using an inverse homomorphism, one can then obtain a
language $K$ from $L$ with $\Psi(K)=\varphi^{-1}(n)$. Thus, $\varphi^{-1}(n)$ is semilinear.

In order to prove the second claim, we present an algorithm to obtain $\nu$ from $\mu$, from which it
will be clear that the size of $\nu$ is linear in the absolute value of $\varphi(\mu)$.
Without loss of generality, let $\varphi(\mu)>0$. The algorithm operates in two phases.

In the first phase, we construct a $\nu\sqsubseteq\mu$ with 
$\varphi(\nu)\ge\varphi(\mu)-m$ such that $|\nu|$ is linear in
$|\varphi(\mu)|$, where $m=\max\{|\varphi(x)| \mid x\in X\}$. In this phase,
we start with $\nu=0$ and successively add elements from $\mu$ to $\nu$ until
$\varphi(\nu)\ge\varphi(\mu)-m$.
As long as we still have $\varphi(\nu)<\varphi(\mu)-m$, it is guaranteed that
we find an $x\in X$ such that $\nu+x\sqsubseteq\mu$ and $\varphi(x)>0$.
Thus, after at most $\varphi(\mu)-m$ steps, we have $\varphi(\nu)\ge\varphi(\mu)-m$ and $|\nu|\le \varphi(\mu)-m$.

Since we stopped after we first had $\varphi(\nu)\ge\varphi(\mu)-m$, we also
have $\varphi(\mu)-m\le\varphi(\nu)\le\varphi(\mu)+m$. In the second phase,
we successively extend $\nu$ such that $\varphi(\nu)$ always stays within
the interval $[\varphi(\mu)-m, \varphi(\mu)+m]$: If $\varphi(\nu)<\varphi(\mu)$, we can find
an $x\in X$ with $\nu+x\sqsubseteq \mu$ and $\varphi(x)>0$ and if $\varphi(\nu)>\varphi(\mu)$,
we can find an $x\in X$ with $\nu+x\sqsubseteq \mu$ and $\varphi(x)<0$.
We do this nondeterministically and can therefore assume that no value
$\varphi(\nu)$ occurs more than once: otherwise, we could have left out the
summands between the two occurrences.

Hence, the values $\varphi(\nu)$ obtained in the course of the second phase are distinct numbers in
$[\varphi(\mu)-m, \varphi(\mu)+m]$. Therefore, there is a computation in which after at most $2m+1$ steps, we have 
$\varphi(\nu)=\varphi(\mu)$. Then clearly $|\nu|\le \varphi(\mu)-m+2m+1$. 
\end{proofqed}

Another fact used in later sections is that languages in $\VA{M}$ are
semilinear if $M\in\PMon$.  This will be used in various constructions, for
instance when the effect of computations (that make use of $M$ as storage) on
the output in a commutative monoid is to be realized by a finite automaton.  We
prove this using a result of van Leeuwen~\cite{vanLeeuwen1974}. He showed that
semilinearity of all languages in a family is inherited by languages that are
algebraic over this family.  A language is called algebraic over a family of
languages if it is generated by a grammar in which each production allows a
non-terminal to be replaced by any word from a language in this family. 
\begin{definition}
Let $\LFamily$ be a family of languages. An \emph{$\LFamily$-grammar} is a
quadruple $G=(V,T,P,S)$ where $V$ and $T\subseteq V$ are alphabets and $S\in
V\setminus T$.  $P$ is a finite set of pairs $(A,M)$ with $A\in V\setminus T$
and $M\subseteq V^*$, $M\in\LFamily$.  In this context, a pair $(A,M)\in P$ will also be denoted by $A\to M$. 
We write $x\step{G} y$ if $x=uAv$ and $y=uwv$ for some
$u,v,w\in V^*$ and $(A,M)\in P$ with $w\in M$. The \emph{language generated by
$G$} is $L(G)=\{w\in T^* \mid S\step{G}^* w\}$. A language $L$ is called
\emph{algebraic over $\LFamily$} if there is an $\LFamily$-grammar $G$ such that
$L=L(G)$.
\end{definition}

We will use the following result by van Leeuwen, which appeared in \cite{vanLeeuwen1974}.
\begin{theorem}[van Leeuwen]\label{thm:algsemilinear}
Let $\LFamily$ be a family of semilinear languages. Then every language that is
algebraic over $\LFamily$ is also semilinear.
\end{theorem}

Note that in \cite{LohreySteinberg2008}, a group $G$ is called \emph{SLI-group} if
every language in $\VAl{G}$ is semilinear (in different terms, however). Thus,
the following recovers the result from \cite{LohreySteinberg2008}, that
the class of SLI-groups is closed under taking the free product.
\begin{lemma}\label{lemma:freeproductalg}
Each $L\in\VAl{M_0*M_1}$ is algebraic over $\VAl{M_0}\cup\VAl{M_1}$.
\end{lemma}
\begin{proofqed}{lemma:freeproductalg}
Let $L=L(A)$ for some valence automaton $A=(Q,X,M_0*M_1,E,q_0,F)$. We assume that
$E=E_0\cup E_1$ with $E_i\subseteq Q\times X^*\times M_i\times Q$ and $F=\{q_f\}$. Let $V_i=\{K_{p,q}^{(i)} \mid p,q\in Q\}$ 
be a new alphabet for $i=0,1$. Furthermore, let $A_{p,q}^{(i)}$ be the automaton
$(Q,X\cup V_{1-i},M_i,E'_i,p,\{q\})$ with $E'_i=E_i\cup \{(r,K_{r,s}^{(1-i)},1,s)\mid r,s\in Q\}$.
Finally, let $A'$ be the automaton $(Q,X\cup V_0\cup V_1,M_0*M_1,E',q_0,F)$ with $E'=E\cup E'_0\cup E'_1$.
Then clearly $L(A)=L(A')\cap X^*$.

We will construct an $\LFamily$-grammar $G=(V,X,P,S)$ such that $L(G)=L(A)$. This will be shown by proving that
for any $w\in V^*$, we have $w\in L(A')$ iff $S\step{G}^* w$.
Let
$V=X\cup V_0\cup V_1$ and $P$ consist of the
productions $K_{p,q}^{(i)}\to L(A_{p,q}^{(i)})$ for $p,q\in Q$ and
$i\in\{0,1\}$. Finally, let $S=K_{q_0,q_f}^{(0)}$. By induction on $n$, one can see that
$S\step{G}^n w$ implies $w\in L(A')$. In particular, $w\in L(G)$ implies that $w$ is accepted by $A'$ without
using the added edges in $E'\setminus E$. Thus, $L(G)\subseteq L(A)$.

Now suppose $w$ is accepted by $A'$, which is witnessed by the edge sequence
$s=u_0v_1u_1\cdots v_nu_n$, where for some $i\in\{0,1\}$, we have $u_0\in
E'^*_i$, $u_j\in E'^+_i$, $v_j\in E'^+_{1-i}$ for each $j\in\{1,\ldots,n\}$.

We shall prove by induction on $n$ that $S\step{G}^* w$. 
For $n=0$, we have $w\in L(A_{q_0,q_f}^{(i)})$ and
thus either $S=K_{q_0,q_f}^{(0)}\step{G} w$ or
$S=K_{q_0,q_f}^{(0)}\step{G}K_{q_0,q_f}^{(1)}\step{G} w$. Therefore, let $n\ge
1$.  Let $\varphi:E'^*\to X^*$, $\psi:E'^*\to M_0*M_1$ be the morphisms with
$\varphi((p,x,m,q))=x$ and $\psi((p,x,m,q))=m$. Then $\varphi(s)=w$ and
$\psi(s)=1$. Since $\psi(u_j)\in M_i$ and $\psi(v_j)\in M_{1-i}$, the
definition of the free product yields that there is a $j$ such that either
$\psi(u_j)=1$ or $\psi(v_j)=1$, where $j>0$ if $u_0=\emptyWord$. Let $x\in E'^+$
be this word $u_j$ or $v_j$.  Note that since $\psi(x)=1$, we have $\varphi(x)
\in L(A_{p,q}^{(k)})$ for $k=i$ or $k=1-i$ (depending on whether $x=u_j$ or
$x=v_j$) and where $p$ and $q$ are the initial and final state of the sequence
$x$, respectively.  We can obtain a sequence $s'$ from $s$ by replacing $x$ with
the edge $(p,K_{p,q}^{(k)},1,q)$.  Clearly, $s'$ is a valid sequence in $A'$
and by induction, we have $S\step{G}^*\varphi(s')$. Since $w$ can be obtained
from $\varphi(s')$ by replacing $K_{p,q}^{(k)}$ with $\varphi(x)\in L(A_{p,q}^{(k)})$, we have
$S\step{G}^*\varphi(s')\step{G}w$.
\end{proofqed}

Combining the latter lemma with van Leeuwen's result and a standard argument
for the preservation of semilinearity when builing the direct product with $\Z$
yields the following.
\begin{lemma}\label{lemma:semilinear}
Let $M\in\PMon$. Then for any language $L\in\VAl{M}$, the set $\Parikh{L}$ is semilinear.
\end{lemma}
\begin{proofqed}{lemma:semilinear}
Since $\VAl{\TrivMon}$ and $\VAl{\B}$ contain only context-free languages, the
lemma holds for $M=\TrivMon$ and $M=\B$. Furthermore, by Lemma
\ref{lemma:freeproductalg} and Theorem \ref{thm:algsemilinear}, we know that if
$\VAl{M}$ contains only semilinear languages then so does $\VAl{M*\B}$.  Thus,
it suffices to show that if every language in $\VAl{M}$ is semilinear then so
is every language in $\VAl{M\times\Z}$. 

Let $L=L(A)$ for a valence automaton $A=(Q,X,M\times\Z,E,q_0,F)$ over
$M\times\Z$.  We assume that for each edge $(p,x,(m,z),q)\in E$, we have $x\in
X\cup \{\emptyWord\}$.  We construct an automaton $A'$ as follows. Let
$A'=(Q,X\times E,M,E',q_0,F)$, where $E'=\{(p,(x,e),m,q) \mid
e=(p,x,(m,z),q)\in E \}$. Furthermore, let $\varphi:(X\times E)^\oplus\to\Z$ be
the morphism with $\varphi((x,(p,y,(m,z),q)))=z$. Moreover, let $\pi: (X\times
E)^\oplus\to X^\oplus$ be the projection map with $\pi((x,e))=x$. Then by the
definition of acceptance, we have
$\Parikh{L(A)}=\pi(\Parikh{L(A')}\cap\varphi^{-1}(0))$. Since $L(A')$ is
semilinear by assumption, the class of semilinear subsets of a free commutative monoid is 
closed under intersection, and $\varphi^{-1}(0)$ is semilinear by Lemma
\ref{lemma:preimages}, $L(A)$ is also semilinear.
\end{proofqed}

\section{Membership Problems}
\label{sec:membership}
In this section, we study decidability and complexity of
the membership problem for valence automata over $\M\Gamma$. 
Specifically, we show in this section that for
certain graphs $\Gamma$, the class $\VAl{\M\Gamma}$ contains undecidable
languages (Lemma \ref{lemma:undecidable}), while for every $\Gamma$,
membership for languages in $\VA{\M\Gamma}$ is (uniformly) decidable. We
present two nondeterministic algorithms, one of them uses linear space 
and one runs in polynomial time (Lemma \ref{lemma:npcs}).

These results serve two purposes. First, for those graphs $\Gamma$ that admit
undecidable languages in $\VAl{\M\Gamma}$, it follows that silent transitions
are indispensable. Second, if we can show that silent transitions can be removed from
valence automata over $\M\Gamma$, the algorithms also apply to languages in $\VAl{\M\Gamma}$.

\newcommand{\rtsclosure}{\overset{*}{\longleftrightarrow}}
\newcommand{\rtclosure}{\overset{*}{\to}}
The algorithms in this section rely on the convergence property of certain
reduction systems.  For more information on reduction systems, see
\cite{Huet1980,BookOtto1993}. 
A \emph{reduction system} is a pair $(S,\to)$ in which $S$ is
a set and $\to$ is a binary relation on $S$.  $(S,\to)$ is said to be
\emph{noetherian} if there is no infinite sequence $s_0,s_1,\ldots$ with
$s_i\to s_{i+1}$ for each $i\in\N$.  We write $\rtsclosure$ ($\rtclosure$) for
the reflexive, transitive, symmetric (reflexive, transitive) closure of $\to$.
$(S,\to)$ has the \emph{Church-Rosser property} if for any $s,t\in S$ with
$s\rtsclosure t$, there is a $u\in S$ with $s\rtclosure u$ and $t\rtclosure u$.
We say that $(S,\to)$ is \emph{confluent}, if for any $s,t,u\in S$ with
$s\rtclosure t$ and $s\rtclosure u$, there is a $v\in S$ with $t\rtclosure v$
and $u\rtclosure v$.  A noetherian and confluent reduction system is called
\emph{convergent}.  Furthermore, $(S,\to)$ is called \emph{locally confluent},
if for any $s,t,u\in S$ with $s\to t$ and $s\to u$, there is a $v\in S$ with
$t\rtclosure v$ and $u\rtclosure v$.  An element $s\in S$ is \emph{irreducible}
if there is no $t\in S$ with $s\to t$.  We say $t\in S$ is a \emph{normal form}
of $s\in S$ if $s\rtclosure t$ and $t$ is irreducible.  It is well-known that a
reduction system is confluent if and only if it has the Church-Rosser property.
Furthermore, a noetherian locally confluent reduction system is already
confluent.

One of the steps in our algorithms will be to check, given a word $w\in
X^*_\Gamma$, whether $w\congruence_\Gamma \emptyWord$. Unfortunately, turning the
Thue system $T_\Gamma$ into a reduction system on words will not yield a
convergent reduction system as the length-preserving rules allow for infinite
reduction sequences. Therefore, we will use reduction systems on traces instead.
For more information on traces, see \cite{DiekertRozenberg1995}.

Let $X$ be an alphabet. An irreflexive symmetric relation $I\subseteq X\times
X$ is called an \emph{independence relation}.  To each such relation, the
corresponding Thue system $T_I=(X, R_I)$ is given as $R_I=\{(ab,ba) \mid (a,b)\in I\}$.  If
$\equiv_I$ denotes the congruence generated by $T_I$, then the monoid
$\T(X,I)=X^*/\mathord{\congruence_I}$ is called \emph{trace monoid}, its elements \emph{traces}.
The equivalence class of $u\in X^*$ is denoted as $[u]_I$ and
since the words in an equivalence class all have the same length, 
$|[u]_I|=|u|$ is well-defined.

In order to efficiently compute using traces, we represent them using
dependence graphs.  Let $X$ be an alphabet and $I\subseteq X\times X$ an
independence relation. To each word $w\in X^*$ we assign a loop-free directed
acyclic vertex-labeled graph, its \emph{dependence graph} $\dep(w)$. If $w=x_1\cdots
x_n$, $x_i\in X$, $1\le i\le n$, then $\dep(w)=(V,E,\ell)$, in which
$E\subseteq V\times V$, has vertex set $V=\{1,\ldots,n\}$ and $(i,j)\in E$ if
and only if $i<j$ and $(x_i, x_j)\notin I$. Furthermore, each vertex $i$ is
labeled with $\ell(i)=x_i\in X$.  It is well-known that for words $u,v\in X^*$,
we have $u\equiv_I v$ if and only if $\dep(u)$ and $\dep(v)$ are isomorphic.
Thus, we will also write $\dep(s)$ for $\dep(u)$ if $s=[u]_I$.

\newcommand{\rcong}[1]{\equiv_{#1|\T}}
\newcommand{\graphtrace}[2]{[#1]_{\Gamma|\T}}
Each (undirected, potentially looped) graph $\Gamma=(V,E)$ gives rise to an independence relation on $X_\Gamma$, namely
\begin{equation}I=\{(x,y) \mid x\in \{a_v, \bar{a}_v\},~ y\in \{a_w, \bar{a}_w\},~ x\ne y,~ \{v,w\}\in E\}. \label{eq:graphindependence}\end{equation}
If $I$ is given by $\Gamma$ in this way, we also write $\rcong{\Gamma}$ for $\equiv_I$ and $\graphtrace{u}{\Gamma}$
instead of $[u]_I$.

In the following, let $I$ be given by $\Gamma=(V,E)$ as in
\eqref{eq:graphindependence}. We will now define a reduction relation $\to$ on
$\T(X_\Gamma, I)$ such that for $u,v\in X^*_\Gamma$
\begin{equation} [u]_\Gamma = [v]_\Gamma ~~~\text{if and only if}~~~ \graphtrace{u}{\Gamma}\rtsclosure\graphtrace{v}{\Gamma}. \label{eq:redequi}\end{equation}
For $s,t\in\T(X_\Gamma, I)$, let $s\to t$ if there are $u_1,u_2\in X^*_\Gamma$
and $v\in V$ such that $s=\graphtrace{u_1a_v\bar{a}_vu_2}{\Gamma}$ and
$t=\graphtrace{u_1u_2}{\Gamma}$.  This definition immediately yields
\eqref{eq:redequi}. Since our algorithms will represent traces as dependence graphs,
we have to restate this relation in terms of the latter. It is not hard to see that
for $s,t\in\T(X_\Gamma, I)$, $s\to t$ if and
only if there are vertices $x,y$ in $\dep(s)$, labeled $a_v$ and $\bar{a}_v$,
respectively, such that
\begin{enumerate}
\item\label{enum:nopathyx} there is no path from $y$ to $x$ and
\item\label{enum:novertexxy} there is no vertex lying on a path from $x$ to $y$
\end{enumerate}
and $\dep(t)$ is obtained from $\dep(s)$ by deleting $x$ and $y$. We will refer to conditions \ref{enum:nopathyx} and \ref{enum:novertexxy} as the \emph{subtrace conditions}.

\begin{figure}
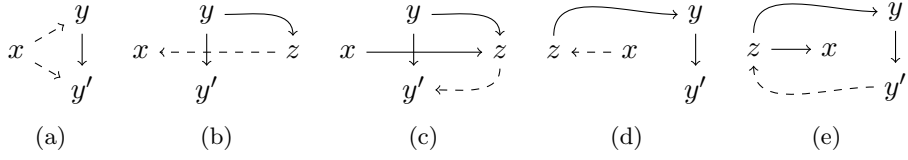

\centering
\subfloat[]{\label{fig:dep:xydep}\depgraphxydep{1}}
\subfloat[]{\label{fig:dep:zrightx}\depgraphzrightx{1}}
\subfloat[]{\label{fig:dep:zrightyp}\depgraphzrightyp{1}}
\subfloat[]{\label{fig:dep:zleftx}\depgraphzleftx{1}}
\subfloat[]{\label{fig:dep:zleftyp}\depgraphzleftyp{1}}
\caption{Possible fragments of the dependence graph of $s$.}
\end{figure}

\begin{lemma}
The reduction system $(\T(X_\Gamma, I), \to)$ is convergent.
\end{lemma}
\begin{proofqed}
Since the system is clearly noetherian, it remains to be shown that
\mbox{$(\T(X_\Gamma, I),\to)$} is locally confluent. Hence, let $x,y,x',y'$ be
vertices in $\dep(s)$ labeled $a_v,\bar{a}_v,a_w,\bar{a}_w$, respectively, 
satisfying the subtrace conditions such that $\dep(t)$ is obtained by deleting $x,y$ and
$\dep(t')$ is obtained by deleting $x',y'$.  If $\{x,y\}=\{x',y'\}$, we are
done.  Furthermore, if $\{x,y\}\cap\{x',y'\}=\emptyset$, deleting $x,y$ from
$\dep(t')$ (or $x',y'$ from $\dep(t)$) yields a $u\in\T(X_\Gamma,I)$ with $t\to
u$ and $t'\to u$.  Therefore, we assume $x=x'$ and $y\ne y'$ (the case $x\ne
x'$, $y=y'$ can be done analogously). This means in particular that $v=w$. Since
$(\bar{a}_v,\bar{a}_v)\notin I$, we can also assume that there is an edge from $y$ to $y'$.

If $(a_v, \bar{a}_v)\notin I$, there are edges $(x,y)$ and $(x,y')$ in
$\dep(s)$ and $y$ violates the second subtrace condition of $x,y'$ (see Figure \ref{fig:dep:xydep}). Hence, we
have $(a_v,\bar{a}_v)\in I$.  We claim that flipping $y$ and $y'$ constitutes
an automorphism of $\dep(s)$, meaning $\dep(t)$ and $\dep(t')$ are isomorphic
and thus $t=t'$.  The former amounts to showing that each vertex $z$ in
$\dep(s)$ has an edge from (to) $y$ iff $z$ has one from (to) $y'$.

If there is an edge from $y$ to $z$, then by the definition of $I$, we also
have an edge between $x$ and $z$. Obeying the first subtrace condition, it has to be directed from $x$ to $z$: Otherwise, there would be a path from $y$ to $x$ (see Figure \ref{fig:dep:zrightx}).  Since $y$ and $y'$ share the same
label, we also have an edge between $y'$ and $z$.  If
this were an edge from $z$ to $y'$, $z$ would lie on a path from $x=x'$ to $y'$ (see Figure \ref{fig:dep:zrightyp}),
violating the second subtrace condition.  Hence, there is an edge from $y'$ to
$z$. 

If there is an edge from $z$ to $y$, then by the definition of $I$, we also have an edge between $x$ and
$z$. By the second subtrace condition, it has to be directed from $z$ to $x$: Otherwise, $z$ would lie on a path from $x$ to $y$ (see Figure \ref{fig:dep:zleftx}).  Since $y$ and $y'$ share the same label, we also have an
edge between $y'$ and $z$.  If this were directed from $y'$
to $z$, then there would be a path from $y'$ to $x=x'$ (see Figure \ref{fig:dep:zleftyp}), violating the first
subtrace condition.  Hence, there is an edge from $z$ to $y'$.

If there is no edge between $y$ and $z$, there is also no edge
between $y'$ and $z$, since $y$ and $y'$ have the same label.
\end{proofqed}

By \eqref{eq:redequi} and since $(\T(X_\Gamma, I), \to)$ is convergent, we have
\begin{equation} [w]_\Gamma=[\emptyWord]_\Gamma ~\text{if and only if}~ \graphtrace{w}{\Gamma}\rtclosure\graphtrace{\emptyWord}{\Gamma}.\label{eq:redequib} \end{equation}
This equivalence is the basis of our algorithms to check for the former condition.

\begin{lemma}\label{lemma:wordp}
There is a deterministic polynomial-time algorithm that, given a word $w\in X_\Gamma^*$, determines whether $[w]_\Gamma=[\emptyWord]_\Gamma$.
\end{lemma}
\begin{proofqednotitle}
By \eqref{eq:redequib}, the condition $[w]_\Gamma=[\emptyWord]_\Gamma$ is equivalent to
$\graphtrace{\emptyWord}{\Gamma}$ being the normal form
of $\graphtrace{w}{\Gamma}$. Therefore, our algorithm computes the normal form
of $\graphtrace{w}{\Gamma}$.  It does so by computing the dependence graph of
$w$ and successively deleting pairs of nodes that satisfy the subtrace
conditions. Finding such a pair can be done in polynomial time and since at
most $|w|/2$ deletions are possible, the normal form is obtained after
polynomial time. In the end, the algorithm checks whether the calculated
dependence graph representing the normal form is empty.
\end{proofqednotitle}

\begin{lemma}\label{lemma:wordcs}
There is a nondeterministic linear-space algorithm that, given a word $w\in
X_\Gamma^*$, determines whether $[w]_\Gamma=[\emptyWord]_\Gamma$.  
\end{lemma}
\begin{proofqednotitle}
Let $\Gamma=(V,E)$. By \eqref{eq:redequib}, we have
$[w]_\Gamma=[\emptyWord]_\Gamma$ if and only if $w$ can be reduced to the empty
word by commuting $a_v$ and $a_w$ for ${v,w}\in E$, commuting $a_v$ and
$\bar{a}_v$ for $\{v\}\in E$ and deleting $a_v\bar{a}_v$ for $v\in V$. 
Therefore, we can clearly construct a monotone grammar for the set of all $w\in X^*_\Gamma$
that have $[w]_\Gamma=[\emptyWord]_\Gamma$.
\end{proofqednotitle}

\begin{lemma}\label{lemma:npcs}
For each $L\in\VA{\M\Gamma}$, the membership problem can be decided by 
\begin{enumerate}
\item a nondeterministic polynomial-time algorithm as well as
\item a nondeterministic linear-space algorithm.
\end{enumerate}
In particular, the languages in $\VA{\M\Gamma}$ are context-sensitive.
\end{lemma}
\begin{proofqed}{lemma:npcs}
In order to decide the membership problem for a word $w$ for a language in
$\VA{\M\Gamma}$, we can guess a run reading $w$. Since there are no
$\lambda$-transitions in the automaton, such a run has length linear in $|w|$.
For this run, we have to check whether the product of the monoid elements on
the edges is the identity element of $\M\Gamma$. By lemmas \ref{lemma:wordp}
and \ref{lemma:wordcs}, this can be done in polynomial time or using linear
space. Hence, the lemma follows.
\end{proofqed}

\begin{lemma}\label{lemma:undecidable}
Let $\Gamma$ be a graph whose underlying loop-free graph is a path on four vertices.
Then $\VAl{\M\Gamma}$ contains an undecidable language.
\end{lemma}
\begin{proofqed}{lemma:undecidable}
Let $\Gamma=(V,E)$ and $\mathring{\Gamma}$ be the graph obtained from $\Gamma$
by adding a loop to every unlooped vertex.  For notational reasons, we assume
that the vertex set of $\mathring{\Gamma}$ is $\mathring{V}=\{\mathring{v}\mid
v\in V\}$.  Lohrey and Steinberg \cite{LohreySteinberg2008} show that there are
rational sets $\mathring{R},\mathring{S}\subseteq\M\mathring{\Gamma}$ over
positive generators such that given a word $w\in\M\mathring{\Gamma}$ over positive generators, it is
undecidable whether $1\in w\mathring{R}\mathring{S}^{-1}$. Note that the
morphism $\varphi:\M\Gamma\to\M\mathring{\Gamma}$ with
$\varphi(a_{v})=a_{\mathring{v}}$ and
$\varphi(\bar{a}_v)=\bar{a}_{\mathring{v}}$ induces an isomorphism between the submonoids
generated by positive generators and between the submonoids generated by the negative generators.
Thus, we find rational sets
$R,S\subseteq\M\Gamma$ over positive generators with $\varphi(R)=\mathring{R}$ and
$\varphi(S)=\mathring{S}$. 

If $w$ is a word over positive generators in $\M\Gamma$, $w=a_1\cdots a_n$, then
we let $\bar{w}=\bar{a}_n\cdots\bar{a}_1$. This is well-defined, for if $a_1\cdots a_n=b_1\cdots b_m$,
for positive generators $a_1,\ldots,a_n,b_1,\ldots,b_m$ then $\varphi(a_1\cdots a_n)=\varphi(b_1\cdots b_m)$
and thus $\varphi(\bar{a}_n\cdots \bar{a}_1)=\varphi(a_1\cdots a_n)^{-1}=\varphi(b_1\cdots b_m)^{-1}=\varphi(\bar{b}_m\cdots\bar{b}_1)$
and therefore $\bar{a}_n\cdots\bar{a}_1=\bar{b}_m\cdots\bar{b}_1$. Note that
$w\bar{w}=1$ for every word $w$ over positive generators. With this definition,
the set $\bar{S}=\{\bar{s}\mid s\in S\}$ is also rational. We claim that for a
word $w\in \M\Gamma$ over positive generators, $1\in wR\bar{S}$ if and only if
$1\in \varphi(w)\mathring{R}\mathring{S}^{-1}$. 

If $1\in \varphi(w)\mathring{R}\mathring{S}^{-1}$, there are
$\mathring{r}\in\mathring{R}$, $\mathring{s}\in\mathring{S}$ with
$1=\varphi(w)\mathring{r}\mathring{s}^{-1}$ and thus
$\mathring{s}=\varphi(w)\mathring{r}$. Thus, we can find $s\in S$ and $r\in R$
with $\varphi(s)=\varphi(w)\varphi(r)$. The injectivity of $\varphi$ on words
over positive generators yields $s=wr$ and thus $1=wr\bar{s}$. Hence $1\in wR\bar{S}$.

If $1\in wR\bar{S}$, we have $1=wr\bar{s}$ for some $r\in R$ and $s\in S$. This
implies $1=\varphi(w)\varphi(r)\varphi(s)^{-1}$ and since $\varphi(r)\in
\mathring{R}$ and $\varphi(s)^{-1}\in \mathring{S}^{-1}$, we have
$1\in\varphi(w)\mathring{R}\mathring{S}^{-1}$.

Thus, given a word $w\in\M\Gamma$ over positive generators, it is undecidable
whether $1\in wR\bar{S}$. Now, we construct a valence automaton over
$\M\Gamma$ that reads a representative of a word $w$ and then
nondeterministically multiplies an element from $R$ and then an element from
$\bar{S}$. It accepts if and only if $1\in wR\bar{S}$.  Therefore, the
automaton accepts an undecidable language.
\end{proofqed}

\section{Rational Sets}
\label{sec:rational}

Here, we present a normal form result for rational subsets of monoids in $\C$.
The first lemma is a simple observation for which we will not provide a proof.
\begin{lemma}\label{lemma:echaract}
Let $s\in M*\B$ and $s=m_1\bar{a}\cdots m_k \bar{a} m a m'_1 \cdots a m'_\ell$. Then we have 
\begin{itemize}
\item $s\in\E{M*\B}$ if and only if
$m_i\in \Li{M}$, $m\in\E{M}$, $m'_j\in\Ri{M}$ for all $1\le i\le k$, $1\le j\le \ell$.
\item $s\in\Li{M*\B}$ if and only if $\ell=0$ and $m_i,m\in\Li{M}$ for $1\le i\le k$.
\item $s\in\Ri{M*\B}$ if and only if $k=0$ and $m'_i,m\in\Ri{M}$ for $1\le i\le \ell$.
\end{itemize}
\end{lemma}

The following lemma states the normal form result. Note that in a valence
automaton over $M$, we can remove all edges labeled with elements outside of
$\E{M}$. This is due to the fact that they cannot be part of a valid
computation. In a valence transducer over $M$ with output in $C$, the edges
carry elements from $X^*\times M\times C$, in which $M$ is used as a storage. Therefore,
of a rational set $S\subseteq M\times C$, we will only be interested in the
part $S\cap(\E{M}\times C)$. 

\todo{More details on how this generalizes other normal form results. Maybe restate lemma in terms
of a language in $X_\Gamma^*$.}

If $A=(Q,M,E,q_0,F)$ is an automaton over $M$ and $B=(Q', M, E', q'_0,
\{q'_f\})$ is an automaton over $M$ with only one final state and $Q\cap
Q'=\emptyset$, then the automaton obtained by \emph{gluing in $B$ between
$p,q\in Q$} is $C=(Q\cup Q', M, E'', q_0, F)$, where
\begin{align*}
E''=E\cup E'\cup \{ (p, 1, q'_0), (q, 1, q'_f) \}.
\end{align*}

\begin{lemma}\label{lemma:ratnormal}
Let $M\in\PMon$ and $C$ be a commutative monoid and $S\subseteq M\times C$ a rational set. Then we have
$S\cap (\E{M}\times C)=\bigcup_{i=1}^n L_iU_iR_i$, in which
\begin{enumerate}
\item[(i)] $L_i\in\RAT{\Li{M}\times C}$, 
\item[(ii)] $U_i\in\RAT{\U{M}\times C}$, and
\item[(iii)] $R_i\in\RAT{\Ri{M}\times C}$
\end{enumerate}
for $1\le i\le n$. Moreover,
\begin{align}
S\cap(\Li{M}\times C)&=\bigcup_{1\le i\le n,~1\in R_i} L_iU_i, \label{eq:ratli}\\
S\cap(\Ri{M}\times C)&=\bigcup_{1\le i\le n,~1\in L_i} U_iR_i. \label{eq:ratri}
\end{align}
\end{lemma}
\begin{proofqed}{lemma:ratnormal}
We proceed by induction. Since the lemma clearly holds for $M=\TrivMon$, we
show that if it holds for $M$, it is also true for $M\times\Z$ and $M*\B$.
Therefore, suppose the lemma holds for $M$.

Let $S\subseteq (M\times\Z)\times C$ be a rational subset. We regard $S$ as a
subset of $M\times(\Z\times C)$.  By induction, we have $S\cap (\E{M}\times
(\Z\times C))=\bigcup_{i=1}^n L_iU_iR_i$ with the properties above. Since
$\X{M\times\Z}=\X{M}\times\Z$ where $\Xo$ is any of the operators
$\Eo$,$\Uo$,$\Lio$,$\Rio$, the sets $L_i, U_i, R_i$ can serve as the desired
decomposition for $(M\times\Z)\times C$.

Let $S\subseteq (M*\B)\times C$ be rational. Then there is an alphabet $X$, a
rational language $L\subseteq X^*$, and a morphism $\varphi: X^*\to
(M*\B)\times C$ with $\varphi(L)=S$. Without loss of generality, we assume that
$X=\{x,\bar{x}\}\cup Y\cup Z$ with $\varphi(x)=a$, $\varphi(\bar{x})=\bar{a}$,
$\varphi(Y)\subseteq M$, $\varphi(Z)\subseteq C$, where $a$ and $\bar{a}$ are
the two generators of $\B$. Let $A$ be an automaton accepting $L$ such that
every edge carries exactly one letter.

As a first step, we will construct an automaton $A'$ that also has
$\varphi(L(A'))=S$ but which has for every element of $S\cap (\E{M*\B}\times
C)$ a representative in $X^*\setminus X^*xX^*\bar{x}X^*$.

Let $A=(Q,X,E,q_0,F)$. For $p,q\in Q$, the language 
\[ K_{p,q} = \{ \pi_Z(w) \mid w\in L_{p,q}(A),~ \varphi(w)\in \{1\}\times C \} \]
is clearly contained in $\VAl{M*\B}$ and is therefore semilinear by Lemma
\ref{lemma:semilinear}. Thus, we can find a finite automaton $A'_{p,q}$ such
that $\Parikh{L(A'_{p,q})}=\Parikh{K_{p,q}}$. Since $C$ is commutative and
$\varphi(Z)\subseteq C$, this also means
$\varphi(L(A'_{p,q}))=\varphi(K_{p,q})$.  The automaton $A'$ is now obtained
from $A$ by gluing $A'_{p,q}$ into $A$ between $p$ and $q$, for each $p,q\in
Q$. 
Since in $A'$ for each path from the initial to the final state, we can find another path
that encodes the same element of $(M*\B)\times C$ and is present in $A$, we have $\varphi(L(A'))=\varphi(L(A))=S$.
However, the glued in automata allow us to encode elements of $S\cap(\E{M*\B}\times C)$
by words of a certain form. Specifically, we claim that
\begin{equation}
S\cap (\E{M*\B}\times C)\subseteq\varphi(L(A')\setminus X^*xX^*\bar{x}X^*) \subseteq S. \label{eq:ecertainform}
\end{equation}
Let $s\in S\cap(\E{M*\B}\times C)$ and $w\in L(A')$ be chosen such that
$\varphi(w)=s$ and $|w|_{X\setminus Z}$ is minimal. Toward a contradiction,
suppose $w\in X^*xX^*\bar{x}X^*$. Then $w=fxg\bar{x}h$ with $f,h\in X^*$, $g\in
(Y\cup Z)^*$.  Since $am\bar{a}\notin\E{M*\B}$ for any $m\in M\setminus\{1\}$,
our assumption $s\in \E{M*\B}\times C$ implies $\varphi(g)\in \{1\}\times C$ and thus
$\varphi(xg\bar{x})\in \{1\}\times C$. By the construction of $A'$, however, 
this means that there is a word $v\in Z^*$ such that $fvh\in L(A')$ and $\varphi(fvh)=\varphi(w)$.
Since $|fvh|_{X\setminus Z}<|w|_{X\setminus Z}$, this contradicts the choice of $w$,
proving \eqref{eq:ecertainform}.

Let $A''=(Q'',X,E'',q''_0,\{q''_f\})$ be a \spelling{} finite automaton accepting $L(A')\setminus X^*xX^*\bar{x}X^*$
with input alphabet $X$. Furthermore,
for each $p,q\in Q''$, let $B^{(-)}_p$, $B^{(0)}_{p,q}$, $B^{(+)}_q$ be
\spelling{} automata satisfying
\begin{align*}
L(B^{(-)}_p)&=L_{q''_0,p}(A'') \cap ((Y\cup Z)^*\bar{x})^* \\
L(B^{(0)}_{p,q})&=L_{p,q}(A'') \cap (Y\cup Z)^* \\
L(B^{(+)}_q)&=L_{q,q''_f}(A'') \cap (x(Y\cup Z)^*)^*
\end{align*}
Then we have
\begin{equation}
 L(A'') = \bigcup_{p,q\in Q''} L(B^{(-)}_p) L(B^{(0)}_{p,q}) L(B^{(+)}_q). \label{eq:intdecomp}
\end{equation}
Since $S\cap (\E{M*\B}\times C)\subseteq \varphi(L(A''))\subseteq S$, we will now modify $A''$ so as to accept only words
whose image lies in $S\cap(\E{M*\B}\times C)$. This will be achieved by exploiting the induction hypothesis for $M\times C$.
From $B^{(-)}_p$, we obtain $\tilde{B}_p^{(-)}$ by removing all edges with letters in $Y\cup Z$ and then between any two states $r$, $s$ gluing in an automaton
that accepts a rational language with image
\[ \varphi(L_{r,s}(B_p^{(-)}) \cap (Y\cup Z)^*)\cap (\Li{M}\times C), \]
which exists by induction. The automaton $\tilde{B}^{(+)}_q$ is obtained from $B^{(+)}_q$ in an analogous way: we remove edges with letters in $Y\cup Z$ and between any two states $r,s$, glue in an automaton that accepts a rational language with image
\[ \varphi(L_{r,s}(B_p^{(+)}) \cap (Y\cup Z)^*)\cap (\Ri{M}\times C). \]
By the induction hypothesis, we can assume that the input alphabet of
$\tilde{B}^{(-)}_p$ and $\tilde{B}^{(+)}_q$ is $\{\bar{x}\}\cup Y^{(-)}\cup
Z$ and $\{x\}\cup Y^{(+)}\cup Z$, respectively, where
$\varphi(Y^{(-)})\subseteq\Li{M}\times C$ and
$\varphi(Y^{(+)})\subseteq\Ri{M}\times C$.
Let $K_p^{(-)}=\varphi(L(\tilde{B}_p^{(-)}))$, $K_q^{(+)}=\varphi(L(\tilde{B}_p^{(+)}))$. Then $K_p^{(-)}\in\RAT{\Li{M*\B}\times C}$,
$K_q^{(+)}\in\RAT{\Ri{M*\B}\times C}$ and
\begin{align*}
K_p^{(-)}&=\varphi(L(B_p^{(-)}))\cap(\Li{M*\B}\times C), \\
K_q^{(+)}&=\varphi(L(B_p^{(+)}))\cap(\Ri{M*\B}\times C).
\end{align*}

Finally, the induction hypothesis provides for each $p,q\in Q''$ rational sets $L_{p,q,i}\in\RAT{\Li{M}\times C}$, $U_{p,q,i}\in\RAT{\U{M}\times C}$, $R_{p,q,i}\in\RAT{\Ri{M}\times C}$ such that
\[ \varphi(L(B_{p,q}^{(0)}))\cap (\E{M}\times C)=\bigcup_{i=1}^{n(p,q)} L_{p,q,i}U_{p,q,i}R_{p,q,i}. \]
We claim that for $\tilde{S}=S\cap (\E{M*\B}\times C)$, we have
\begin{equation}
\tilde{S}=\bigcup_{\underset{1\le i\le n(p,q)}{p,q\in Q''}}  \left[ K_p^{(-)} L_{p,q,i}\right] U_{p,q,i} \left[R_{p,q,i} K_q^{(+)}\right].\label{eq:ratdecomp} 
\end{equation}
Thus, let $s\in \tilde{S}$. By \eqref{eq:ecertainform}, the definition of $A''$, and \eqref{eq:intdecomp}, we can write
$s=\varphi(w)$ with 
\[ w = u_1\bar{x}\cdots u_k\bar{x} v xw_1 \cdots x w_\ell \]
and $u_i,v,w_j\in (Y\cup Z)^*$ for $1\le i\le k$, $1\le j\le \ell$ and
$u_1\bar{x}\cdots u_k\bar{x}\in L(B_p^{(-)})$, $v\in L(B^{(0)}_{p,q})$,
$xw_1\cdots xw_\ell\in L(B^{(+)}_q)$ for some $p,q\in Q''$.
By Lemma \ref{lemma:echaract}, the fact that $s\in \E{M*\B}\times C$ implies that 
$\varphi(u_i)\in \Li{M}\times C$, $\varphi(v)\in\E{M}\times C$, and $\varphi(w_j)\in\Ri{M}\times C$ for $1\le i\le k$, $1\le j\le \ell$.
In particular, there are words $\tilde{u}_i\in (Y^{(-)}\cup Z)^*, \tilde{w}_j\in (Y^{(+)}\cup Z)^*$ such that
$\varphi(\tilde{u}_i)=\varphi(u_i)$, $\varphi(\tilde{w}_j)=\varphi(w_j)$ and
\[ \tilde{u}_1\bar{x}\cdots \tilde{u}_k\bar{x}\in L(\tilde{B}^{(-)}_p),~~x\tilde{w}_1\cdots x\tilde{w}_\ell\in L(\tilde{B}^{(+)}_q) \]
and $\varphi(v)\in L_{p,q,i}U_{p,q,i}R_{p,q,i}$ for some $1\le i\le n(p,q)$, proving ``$\subseteq$'' of \eqref{eq:ratdecomp}.

For the inclusion ``$\supseteq$'', note that by \eqref{eq:ecertainform} and the definition of the sets on the right, the right side is contained in $S$. Moreover, 
by Lemma \ref{lemma:echaract}, the right side is also contained in $\E{M*\B}\times C$. This proves \eqref{eq:ratdecomp}.

We will show that \eqref{eq:ratdecomp} is the desired decomposition of $\tilde{S}$. 
We have already established that
\begin{enumerate}
\item[(i)] $K_p^{(-)} L_{p,q,i}\in \RAT{\Li{M*\B}\times C}$,  
\item[(ii)] $U_{p,q,i}\in\RAT{\U{M}\times C}\subseteq\RAT{\U{M*\B}\times C}$, and
\item[(iii)] $R_{p,q,i}K^{(+)}_q\in \RAT{\Ri{M*\B}\times C}$. 
\end{enumerate}
Therefore, it remains to be shown that \eqref{eq:ratli} and \eqref{eq:ratri} are satisfied.
We only prove \eqref{eq:ratli}, the proof for \eqref{eq:ratri} can be done analogously.
The inclusion ``$\supseteq$'' is immediately clear in each case.
Thus, suppose $s\in S\cap(\Li{M*\B}\times C)$. Then there is a word $w\in L(A'')$ with $s=\varphi(w)$ and
\[ w = u_1\bar{x}\cdots u_k\bar{x} v xw_1 \cdots x w_\ell \]
and $u_i,v,w_j\in (Y\cup Z)^*$ for $1\le i\le k$, $1\le j\le \ell$ and
$u_1\bar{x}\cdots u_k\bar{x}\in L(B_p^{(-)})$, $v\in L(B^{(0)}_{p,q})$,
$xw_1\cdots xw_\ell\in L(B^{(+)}_q)$ for some $p,q\in Q''$.
By Lemma \ref{lemma:echaract}, the fact that $s\in\Ri{M*\B}\times C$ implies
$\varphi(u_i),\varphi(v)\in\Li{M}\times C$ for $1\le i\le k$ and $\ell=0$.
In particular, there are words $\tilde{u}_i\in (Y^{(-)}\cup Z)^*$ such that
$\varphi(\tilde{u}_i)=\varphi(u_i)$ and
\[ \tilde{u}_1\bar{x}\cdots \tilde{u}_k\bar{x}\in L(\tilde{B}^{(-)}_p),~~~ \emptyWord\in L(\tilde{B}^{(+)}_q) \]
and $\varphi(v)\in L_{p,q,i}U_{p,q,i}$ and $1\in R_{p,q,i}$ for some $1\le i\le n(p,q)$ by induction.
Therefore, $1\in R_{p,q,i}K_q^{(+)}$ and $s\in K_p^{(-)}L_{p,q,i}U_{p,q,i}$. This proves the remaining inclusion ``$\subseteq$'' of \eqref{eq:ratli}.
\end{proofqed}

If we regroup the factors in \eqref{eq:ratdecomp}, we obtain
\begin{equation*}
\tilde{S}=\bigcup_{\underset{1\le i\le n(p,q)}{p,q\in Q''}}  K_p^{(-)} \left[ L_{p,q,i} U_{p,q,i} R_{p,q,i}\right] K_q^{(+)},
\end{equation*}
which implies the following corollary. Note, however, that it can also be deduced from Lemma \ref{lemma:ratnormal}.
\begin{corollary}\label{cor:ratnormal}
Let $M\in\C$ and $S$ be a rational subset of $(M*\B)\times C$. Then there is an alphabet $X=\{x,\bar{x}\}\cup Y\cup Z$, a morphism
$\varphi:X^*\to (M*\B)\times C$ with $\varphi(x)=a$, $\varphi(\bar{x})=\bar{a}$, $\varphi(Y)\subseteq M$, $\varphi(Z)\subseteq C$, a number $n\in\N$, and rational languages
\[ L_i\subseteq ((Y\cup Z)^*\bar{x})^*, ~~ J_i\subseteq (Y\cup Z)^*,~~R_i\subseteq (x(Y\cup Z)^*)^* \]
such that 
\[S\cap (\E{M*\B}\times C)=\bigcup_{i=1}^n \varphi(L_i)\varphi(J_i)\varphi(R_i).\]
\end{corollary}

\section{Silent Transitions}
\label{sec:silent}

\newcommand{\PosR}[2]{\bigtriangleup_{#1,#2}^\circ}
\newcommand{\NegR}[2]{\bigtriangleup_{#1,#2}^\bullet}
\newcommand{\State}[1]{\Box_{#1}}
\newcommand{\Between}{\Box}
\newcommand{\Op}[1]{\mathsf{#1}}
\newcommand{\LoopedPath}{\ensuremath{P_4^\circ}}

In this section, we use the facts established in earlier sections to prove the main results.

\begin{lemma}\label{lemma:graph}
Let $\Gamma$ be a graph such that
\begin{itemize}
\item between any two looped vertices, there is an edge, and
\item between any two unlooped vertices, there is no edge, and
\item $\Gamma$ does not contain \loopedpath{1} as an induced subgraph. 
\end{itemize}
Then $\M\Gamma$ is in $\C$.
\end{lemma}
\begin{proofqed}{lemma:graph}
We proceed by induction and thus assume that $\M(\Gamma\setminus\{x\})\in \C$ for any vertex $x$.
Let $\Gamma=(V,E)$ and write $V=L\cup U$, where $L$ is the set of looped
vertices and $U$ is the set of unlooped vertices. For every $x\in L$, let
$\nu(x)=N(x)\cap U$, i.e. the set of unlooped neighbors of $x$. We write
$x\le y$ for $x,y\in L$ if $\nu(x)\subseteq\nu(y)$. Clearly, $\le$ is a reflexive, transitive order on $L$.

If there were $x,y\in L$ such that $\nu(x)$ and $\nu(y)$ are incomparable,
there would be vertices $u,v\in U$ with $u\in \nu(x)\setminus\nu(y)$ and
$v\in\nu(y)\setminus\nu(x)$. Thus, the vertices $u,x,y,v$ induce the subgraph
\loopedpath{0.5}, contradicting the hypothesis. Hence, $\le$ is a total order
and has a greatest element $g\in L$.
\begin{itemize}
\item If $\nu(g)=U$, then $g$ is adjacent to every vertex in $\Gamma$ and thus
$\M\Gamma\cong \M(\Gamma\setminus\{g\})\times\Z$.
\item If $\nu(g)\subsetneq U$, then there is an isolated vertex $u\in U\setminus \nu(g)$.
Hence, we have $\M\Gamma\cong\M(\Gamma\setminus\{u\})*\B$.
\end{itemize}
\end{proofqed}

We will prove Theorem \ref{theorem:firstmain} by showing that that
$\VA{M}=\VAl{M}$ for every $M\in\C$. This will be done using an induction with
respect to the definition of $\C$. In order for this induction to work, we need
to strengthen the induction hypothesis. The latter will state that for any $M\in\C$ and any
commutative monoid $C$, we can transform a valence transducer over $M$ with output in $C$
into another one that has no $\emptyWord$-transitions but is allowed to output a semilinear
set of elements in each step. Formally, we will show that each $M\in\C$ is \emph{\lelim{}}.
\begin{definition}
Let $C$ be a commutative monoid and $T\subseteq X^*\times\SL{C}$ be a transduction.
Then $\SLMap{T}\subseteq X^*\times C$ is defined as
\[\SLMap{T}=\{(w,c)\in X^*\times C \mid \exists (w,S)\in T: c\in S \}. \]
For a class $\F$ of transductions, $\SLMap{\F}$ is the class of all $\SLMap{T}$ with $T\in\F$.

A monoid $M$ is called \emph{\lelim{}} if for any commutative monoid $C$, we have
$\VTl{M}{C}=\SLMap{\VT{M}{\SL{C}}}$.
\end{definition}
Note that the inclusion $\SLMap{\VT{M}{\SL{C}}}\subseteq\VTl{M}{C}$ holds for
any $M$ and $C$.  Here, in order to have equality, it is necessary to
grant the $\emptyWord$-free transducer the output of semilinear sets, since valence
transducers without $\emptyWord$-transitions and with output in $C$ can only
output finitely many elements per input word. With $\emptyWord$-transitions,
however, a valence transducer can output an infinite set for one input word. 

By choosing the trivial monoid for $C$, we can see that for every \lelim{}
monoid $M$, we have $\VA{M}=\VAl{M}$.  Indeed, given a valence automaton $A$
over $M$, add an output of $1$ to each edge and transform the resulting valence
transducer into \anEmpty{} $\emptyWord$-free one with output in $\SL{\TrivMon}$. The
latter can then clearly be turned into a valence automaton for the language
accepted by $A$.  Thus, we have the following lemma.
\begin{lemma}\label{lemma:stronger}
If $M$ is \lelim{}, then $\VA{M}=\VAl{M}$.
\end{lemma}

\begin{definition}
A \emph{rationally labeled valence transducer} over $M$ with output in $C$ is
an automaton over $X^*\times\RAT{M\times C}$.  For $A=(Q,X^*\times\RAT{M\times
C},E,q_0,F)$, we also write $A=(Q,X,M,C,E,q_0,F)$. The \emph{transduction
performed by $A$} is 
\[ T(A)=\{ (w,c)\in X^*\times C \mid \exists q\in F: (q_0,(\emptyWord,\{1\}))\step{A}^* (q,(w,S)),~(1,c)\in S\}. \]
$A$ is called \emph{\spelling{}} if $E\subseteq Q\times X\times\RAT{M\times
C}\times Q$, i.e., if it reads exactly one letter in each transition.
\end{definition}
The definition of $T(A)$ for rationally labeled valence transducers $A$
means that $A$ behaves as if instead of an edge $(p, (w, S), q)$, $S\in\RAT{M\times C}$, 
it had an edge $(p, w, m, c, q)$ for each $(m,c)\in S$. Therefore, in slight
abuse of terminology, we will also say that
\[ q_0 \xrightarrow{(x_1, m_1, c_1)} q_1\rightarrow \cdots\rightarrow q_{n-1}\xrightarrow{(x_n,m_n,c_n)}q_n \]
is a computation in $A$ when there are edges $(q_{i-1}, (x_i, S_i), q_i)\in E$ such
that $(m_i,c_i)\in S_i$ for $1\le i\le n$.

\begin{lemma}\label{lemma:rationallabels}
For each valence transducer $A$ over $M$ with output in $C$, there is a
\spelling{} rationally labeled valence transducer $A'$ with $T(A')=T(A)$.
\end{lemma}
\begin{proofqednotitle}
Let $A=(Q,X,M,C,E,q_0,F)$.
We obtain the $\emptyWord$-free rationally labeled valence transducer
$A'=(Q,X,M,C,E',q_0,F)$ as follows.
We introduce one edge $(p,(x,S),q)$ for every triple $(p,x,q)\in Q\times
X\times Q$ such that $S\subseteq M\times C$ is the rational set of elements
spelled by paths in $A$ that start in $p$, go along a number of
$\emptyWord$-edges, then pass through an edge labeled $x$ and then again go along
a number of $\emptyWord$-edges and stop in $q$. Then clearly $T(A')=T(A)$.
\end{proofqednotitle}

\begin{lemma}\label{lemma:b}
$\B$ is \lelim{}.
\end{lemma}
\begin{proofqed}{lemma:b}
Let $T\in\VTl{\B}{C}$. By Lemma \ref{lemma:rationallabels}, we can assume that $T=T(A)$
for a rationally labeled valence transducer $A=(Q,X,\B,C,E,q_0,F)$ over $\B$ with output in $C$.

\newcommand{\SSpan}[1]{\{#1\}^\oplus}
\newcommand{\Span}[1]{#1^\oplus}
By Lemma \ref{lemma:ratnormal}, we can assume that every edge in $A$ has the
form $(p,x,LR,q)$, with $L\in\RAT{\SSpan{\bar{a}}\times C}$ and
$R\in\RAT{\SSpan{a}\times C}$.  Furthermore, we can assume that edges starting
in the initial state $q_0$ are of the form $(q_0,x,R,p)$ and, analogously,
edges ending in a final state $q\in F$ are of the form $(p,x,L,q)$, $p\in Q$,
$L\in\RAT{\SSpan{\bar{a}}\times C}$ and $R\in\RAT{\SSpan{a}\times C}$. Thus, we
can construct an equivalent transducer $A'=(Q',X,\B,C,E',q_0,F')$ each edge of which simulates the
$R$-part of one edge of $A$ and then the $L$-part of another edge of $A$. Hence, 
in $A'$, every edge is of the form $(p,x,RL,q)$ with $p,q\in Q'$,
$R\in\RAT{\SSpan{a}\times C}$, and $L\in\RAT{\SSpan{\bar{a}}\times C}$.

Since $\SSpan{a}\times C$ and $\SSpan{\bar{a}}\times C$ are commutative, all
such $R$ and $L$ are semilinear sets and we can even assume that every
edge is of the form $(p,x,\Span{R}(m,c)\Span{L},q)$, in which $(m,c)\in\B\times
C$ and $R$ and $L$ are finite subsets of $\SSpan{a}\times C$ and
$\SSpan{\bar{a}}\times C$, respectively.

Now the first crucial observation is that if we allow the transducer to apply
elements of $\SSpan{a}\times C$ that, in an edge
$(p,x,\Span{R}(m,c)\Span{L},q)$ traversed earlier, were contained in $R$, we do
not increase the set of accepted pairs in $X^*\times C$. This is due to the
fact that if the counter realized by $\B$ does not go below zero in this new
computation, it will certainly not go below zero if we add the value at hand in an
earlier step.  Thus, any computation in the new transducer can be transformed
into one in the old transducer. Furthermore, the commutativity of $C$
guarantees that the output is invariant under this transformation. Analogously,
if we allow the transducer to apply elements from $\SSpan{\bar{a}}\times C$, as
long as it ensures that in some edge $(p,x,\Span{R}(m,c)\Span{L},q)$ traversed later, they are
contained in $L$, we do not change the accepted set of pairs either.

Therefore, we construct a rationally labeled transducer $A''$ from $A'$.  In
its state, $A''$ stores a state of $A'$ and two sets: a finite set
$\tilde{R}\subseteq\SSpan{a}\times C$ and a finite set
$\tilde{L}\subseteq\SSpan{\bar{a}}\times C$.  $\tilde{R}$ always contains all
those elements of $\SSpan{a}\times C$ that have occurred in sets $R$ so far, and
$\tilde{L}$ are elements of $\SSpan{\bar{a}}\times C$ that still have to be
encountered in sets $L$ in the future. Then for every edge
$(p,x,\Span{R}(m,c)\Span{L},q)$ in $A'$, we have an edge labeled
$(x,\Span{(R\cup \tilde{R})}(m,c)\Span{(L\cup\tilde{L})})$.  $A''$ will then add
the elements of $R$ to its set $\tilde{R}$ and nondeterministically remove some
elements of $L$ from $\tilde{L}$ (they can only be removed if this is their last
occurrence; otherwise, we might need them in $\tilde{L}$ later). The final
state will then make sure that $\tilde{L}$ is empty and $A''$ has thus only
applied elements early that would later appear. In the initial state, both sets $\tilde{R}$
and $\tilde{L}$ are empty and then $\tilde{L}$ is filled nondeterministically.

We have constructed $A''$ to have the following property. For every computation
\[q_0 \xrightarrow{(x_1, \Span{R_1}(m_1,c_1)\Span{L_1})} q_1 \cdots q_{n-1}\xrightarrow{(x_n, \Span{R_n}(m_n,c_n)\Span{L_n})} q_n, \]
we have 
\[R_1\subseteq R_2\subseteq\cdots\subseteq R_n ~~\text{and}~~L_1\supseteq L_2\supseteq\cdots\supseteq L_n. \]

The essential idea of the proof is that in $A''$, we can accept any pair of
$X^*\times C$ by a computation such that the element in the $R^\oplus$-part and
the element in the $L^\oplus$-part in each edge differ in length (as measured
by the number of $a$ and $\bar{a}$) only by a bounded number. If they differ by
more than the bound, either some part of the $R^\oplus$-part or some part of the
$L^\oplus$-part can be postponed or applied earlier, respectively. 

If then we know that these lengths differ only by a bounded number, we will see
that for each occurring difference (between the lengths), the set of possible
outputs is semilinear.  Thus, we only have to output this semilinear set and
add this difference.

The valence transducer $\hat{A}$ is obtained from $A''$ as follows. Let
$e=(p,x,\Span{R}(a^k\bar{a}^n, c)\Span{L}, q)$ be an edge in $A''$.  Let $Y$ and
$Z$ be alphabets in bijection with $R$ and $L$, respectively, and let
$\varphi:(Y\cup Z)^\oplus\to\B\times C$ be the morphism extending these
bijections. Furthermore, if $\kappa:\B\to\Z$ is the morphism with $\kappa(a)=1$
and $\kappa(\bar{a})=-1$, let $\psi:(Y\cup Z)^\oplus \to\Z$ be defined by
$\psi(\mu)=\kappa(\pi_1(\varphi(\mu)))$.  The set
$C_i=\pi_2(\varphi(\psi^{-1}(i)))\subseteq C$ now contains all outputs $c_1c_2\in
C$ such that there are $(a^t, c_1)\in R^\oplus$ and $(\bar{a}^u,c_2)\in L^\oplus$
with $t-u=i$. Moreover, by Lemma \ref{lemma:preimages}, the set $C_i$ is
semilinear. Let 
\begin{equation} b=\min \{-1, \psi(z)+n-k \mid z\in Z\},~~~B=\max\{1, \psi(y)+n-k \mid y\in Y\}.\label{eq:minmax} \end{equation}
$\hat{A}$ has the same set of states as $A''$.
To simulate the edge $e$, we introduce for each $i\in\N$ with $b< i< B$ the edge
\begin{align} 
(p, x, a^{k+i}\bar{a}^n, cC_i, q) &~~~~\text{if $i\ge 0$,} \label{eq:posedge} \\
(p, x, a^k\bar{a}^{n-i}, cC_i, q) &~~~~\text{if $i<0$.}    \label{eq:negedge}
\end{align}
Initial state and final states remain unaltered. We claim that $\SLMap{T(\hat{A})}=T(A'')$.
By the construction, it is clear that $\SLMap{T(\hat{A})}\subseteq T(A'')$. Now
consider a computation in $A''$ with steps $p\xrightarrow{(x,r(a^k\bar{a}^n,
c)\ell)}q$ for edges $(p,x,R^\oplus(a^k\bar{a}^n, c)L^\oplus, q)$.  Define
$\varphi$, $\kappa$, $\psi$, $m$, $M$ as above. Let $r=\varphi(\mu)$ and
$\ell=\varphi(\nu)$, $\mu\in Y^\oplus$, $\nu\in Z^\oplus$. 

Suppose there is a $y$ in $\mu$ such that 
\begin{equation} \psi(\mu-y)+k-n+\psi(\nu)\ge 0, \label{eq:moveright}\end{equation}
that is, the counter stays above zero until the end of the step, even if we do
not add $y$. Then the counter will also stay above zero if we postpone the
application of $\varphi(y)$ until the beginning of the next step.  By
construction, $A''$ allows us to do so. Note that we cannot be in the last step
of the computation, since this would leave a positive value on the
counter.  Analogously, suppose there is a $z$ in $\nu$ such that 
\begin{equation} -\psi(\nu-z)+n-k-\psi(\mu)\ge 0, \label{eq:moveleft}\end{equation}
that is, when starting from the right (and interpreting $\bar{a}$ as increment
and $a$ as decrement), the counter does not drop below zero until the beginning
of the step, even if we do not apply $\varphi(z)$.  Then we can apply $\varphi(z)$ earlier
in the computation. Again, note that this cannot happen in the first step, since this would
mean the computation starts by subtracting from the counter.

We transform the computation in the following way. Whenever in some step,
\eqref{eq:moveright} is satisfied, we move $\varphi(y)$ to the right (i.e., we
postpone the application of $\varphi(y)$).  Symmetrically, whenever in some step,
\eqref{eq:moveleft} is fulfilled, we move $\varphi(z)$ to the left (i.e., we
apply $\varphi(z)$ earlier). We repeat this and since the computation is finite, this
process will terminate and we are left with a valid equivalent computation in which
\eqref{eq:moveright} and \eqref{eq:moveleft} do not occur.

The equations \eqref{eq:moveright} and \eqref{eq:moveleft} are equivalent to
\begin{eqnarray*}
\psi(\mu)+\psi(\nu)&\ge& \psi(y)+n-k, \\
\psi(\mu)+\psi(\nu)&\le& \psi(z)+n-k.
\end{eqnarray*}
Since these are not satisfied, we have
\begin{eqnarray}
\psi(\mu)+\psi(\nu)&<& \psi(y)+n-k ~~~\text{for each $y$ in $\mu$},\label{eq:nomoveright} \\
\psi(\mu)+\psi(\nu)&>& \psi(z)+n-k ~~~\text{for each $z$ in $\nu$}\label{eq:nomoveleft}
\end{eqnarray}
and thus
\[ b < \psi(\mu)+\psi(\nu) < B. \]
Note that these inequalities follow from \eqref{eq:nomoveleft} and \eqref{eq:nomoveright}, respectively, if $\nu\ne 0$ and $\mu\ne 0$.
In case $\mu=0$ or $\nu=0$, they still hold because then $\psi(\mu)+\psi(\nu)$ is $\le 0$ or $\ge 0$, respectively, and
$b<0$ and $0<B$. This means, however, that each step has a counterpart in the edges \eqref{eq:posedge} and \eqref{eq:negedge}.
Therefore, $\SLMap{T(\hat{A})}=T(A'')$.
\end{proofqed}

\begin{lemma}\label{lemma:z}
Suppose $M\in\C$ is \lelim{}. Then $M\times\Z$ is \lelim{} as well.
\end{lemma}
\begin{proofqed}{lemma:z}
In order to simplify notation, we write the operation of $C$ with $+$.
Let $T\in\VTl{M\times\Z}{C}$ and let $A=(Q,X,M\times\Z,C,E,q_0,F)$ be a
transducer for $T$.  By letting $E'=\{(p,x,m,(z,c),q) \mid (p,x,(m,z),c,q)\in
E\}$, we get a transducer $A'=(Q,X,M,\Z\times C,E',q_0,F)$.  Then we have
$(w,c)\in T$ if and only if there is a $(w,(0,c))\in T(A')$.  By the
hypothesis, there is \anEmpty{} $\emptyWord$-free valence transducer $A''$ over $M$ with
output in $\SL{\Z\times C}$ such that $\SLMap{T(A'')}=T(A')$.

In $A''$, every edge is of the form $(p,x,m,S,q)$, where $S\subseteq\Z\times C$
is semilinear.  Thus, we can assume that every edge is of the form
$(p,x,m,(\ell,c)+S^\oplus,q)$, where $S\subseteq\Z\times C$ is finite.  Since
$\Z\times C$ is commutative, we do not change the transduction if we output
elements $s\in\Z\times C$ that occur in some $S$ in a step anywhere else in the
computation. Therefore, we can transform $A''$ so as to make it guess the set
$\tilde{S}$ of all $s\in\Z\times C$ that will occur in an $S$ somewhere in the
computation. It uses its finite control to guarantee that the computation is
only accepting if all elements of $\tilde{S}$  actually occur. In every step,
it allows the output of every element of $\tilde{S}^\oplus$. Thus, in the
resulting transducer $A'''$, we have that in any computation, the set $S$ in
steps $p\xrightarrow{(x,m,(\ell,c)+S^\oplus)}q$ does not change throughout the
computation.

Our new transducer $\hat{A}$ has the same set of states as $A'''$ and the edges
are defined as follows.  For the edge $(p,x,m,(\ell,c)+S^\oplus,q)$ in $A'''$,
let $Y$ be an alphabet in bijection with $S$ and let $\varphi:Y^\oplus \to
\Z\times C$ be the morphism extending this bijection. Let $k\in\N$ be the
constant provided by Lemma \ref{lemma:preimages} for the map $\psi:Y^\oplus\to\Z$,
$\psi(\mu)=\pi_1(\varphi(\mu))$. We introduce an edge
\[ (p,x,(m,\ell+\psi(\nu)), c+\pi_2(\varphi(\nu+\ker\psi)), q) \]
for every $\nu\in Y^\oplus$ with $|\nu|\le k\cdot B$, where $B$ is the maximum
over all values $|\ell|$ for edges $(p',x',m',(\ell,c')+S^\oplus,q')$ in $A'''$.  Initial
and final states remain unaltered.  Note that by Lemma \ref{lemma:preimages},
the set $c+\pi_2(\varphi(\nu+\ker\psi))\subseteq C$ is semilinear.  Observe
that each of these edges chooses an element of $S^\oplus$, namely a
$\mu\in\nu+\ker\psi$, and adds $\ell+\psi(\nu)=\ell+\psi(\mu)$ to the
$\Z$-component of the storage and outputs $c+\pi_2(\varphi(\mu))$.  Thus, it
simulates a step in $A'''$. Therefore, if $(w,c)\in \SLMap{T(\hat{A})}$, then
$(w,(0,c))\in \SLMap{T(A''')}$ and thus $(w,c)\in T$.

It remains to be shown that $(w,(0,c))\in \SLMap{T(A''')}$ implies $(w,c)\in\SLMap{T(\hat{A})}$.
Therefore, Let
\[q_0 \xrightarrow{(x_1,m_1,(\ell_1,c_1)+s_1)} q_1 \cdots q_{n-1} \xrightarrow{(x_n,m_n,(\ell_n,c_n)+s_n)} q_n \]
be a computation in $A'''$ that witnesses $(w,(0,c))\in \SLMap{T(A''')}$. Let
$s_i\in S^\oplus$ for $1\le i\le n$ and define $Y,\varphi,\psi,k,B$ as above.
Let $s_i=\varphi(\mu_i)$, $\mu_i\in Y^\oplus$. Since the computation accepts
$(w, (0,c))$, we have $\psi(\mu_1+\cdots\mu_n)+\ell_1+\cdots\ell_n=0$ and for
$\mu=\mu_1+\cdots+\mu_n$ we have thus 
\[|\psi(\mu)|=|\psi(\mu_1+\cdots+\mu_n)|=|\ell_1+\cdots+\ell_n|\le n\cdot B. \]
Lemma \ref{lemma:preimages} now yields a $\nu\sqsubseteq\mu$ with
$\mu\in\nu+\ker\psi$ and $|\nu|\le knB$.  This means that we can write
$\nu=\nu_1+\cdots+\nu_n$ such that $|\nu_i|\le kB$ for $1\le i\le n$.  Since
$\mu\in\nu+\ker\psi$, we have $s_i\in\varphi(\nu+\ker\psi)$ and
\[ \ell_1+\psi(\nu_1)+\cdots+\ell_n+\psi(\nu_n)=\ell_1+\cdots+\ell_n+\psi(\mu)=0. \]
Thus, using the edges
\[ q_0\xrightarrow{(x_1,(m_1,\ell_1+\psi(\nu_1)), c_1+\pi_2(\varphi(\nu_1+\ker\psi)))} q_1\cdots q_{n-1}\xrightarrow{(x_n,(m_n,\ell_n+\psi(\nu_n)), c_n+\pi_2(\varphi(\nu_n+\ker\psi)))} q_n,\]
we have $(w,c)\in\SLMap{T(\hat{A})}$.
\end{proofqed}

\begin{lemma}\label{lemma:inclusion}
If $\varphi:M\to N$ is a morphism with $\varphi^{-1}(1)=\{1\}$, then $\VT{M}{C}\subseteq \VT{N}{C}$ for any monoid $C$.
\end{lemma}
\begin{proofqednotitle}
Take a valence transducer over $M$ with output in $C$ and replace each edge
$(p,x,m,c,q)$ by $(p,x,\varphi(m),c,q)$.  This yields a valence transducer over
$N$ that performs the same transduction.
\end{proofqednotitle}

\begin{lemma}\label{lemma:stack}
Suppose $M\in\C$ is non-trivial and \lelim{}. Then $M*\B$ is \lelim{} as well.
\end{lemma}
\begin{proofqed}{lemma:stack}
By Lemma \ref{lemma:rationallabels}, in order to show $T\in
\SLMap{\VT{M*\B}{\SL{C}}}$ for any given $T\in\VTl{M*\B}{C}$, we can assume that
$T=T(A)$ for a rationally labeled valence transducer $A$ over $M*\B$ with output in
$C$. Without loss of generality, we can assume that in $A=(Q,X,M*\B,C,E,q_0,F)$,
we have $E\subseteq Q\times X\times \RAT{(M*\B)\times C}\times Q$ and $F=\{q_f\}$.

First, we claim that $\VT{M*\B^{(n)}}{C}=\VT{M*\B}{C}$ for any commutative $C$.
In fact, since $M\in\C$ is non-trivial, it contains an element $b\in \Ri{M}$
such that $b^i\ne b^j$ for $i\ne j$, $i,j\in\N$. Let $b\bar{b}=1$ and let
$a_1,\overline{a_1},\ldots,a_n,\overline{a_n}$ be the generators of the factors
$\B$ in $M*\B^{(n)}$, respectively. Then the map $\varphi:M*\B^{(n)}\to M*\B$,
with $\varphi(a_i)=ab^ia$, $\varphi(\overline{a_i})=\bar{a}\bar{b}^i\bar{a}$,
$\varphi(m)=m$, for $1\le i\le n$ and $m\in M$, clearly satisfies $\varphi^{-1}(1)=\{1\}$.
Thus, by Lemma \ref{lemma:inclusion}, we have
$\VT{M*\B^{(n)}}{C}=\VT{M*\B}{C}$ and it will suffice to show
\[ T(A)\in \SLMap{\VT{M*\B^{(n)}}{\SL{C}}} \]
for some $n\in\N$. 

By Corollary \ref{cor:ratnormal}, we can assume that for every edge $(p,(x,S),q)\in E$, there is an alphabet $X'=\{x,\bar{x}\}\cup Y\cup Z$,
a morphism $\varphi:X'^*\to (M*\B)\times C$ with $\varphi(x)=a$, $\varphi(\bar{x})=\bar{a}$, $\varphi(Y)\subseteq M$, $\varphi(Z)\subseteq C$, and rational languages
\begin{equation}\label{eq:xdist}
L\subseteq ((Y\cup Z)^*\bar{x})^*, ~~ J\subseteq (Y\cup Z)^*,~~R\subseteq (x(Y\cup Z)^*)^* 
\end{equation}
such that $S\cap (\E{M*\B}\times C)=\varphi(L)\varphi(J)\varphi(R)$. Indeed:
for those edges where the union provided by Corollary \ref{cor:ratnormal}
ranges over more than one set, we can introduce new edges. Without loss of
generality, we assume that the alphabets $X',Y,Z$ are the same for all edges
$(p,(x,S),q)$.  If we replace the edge $(p,(x,S),q)$ by
$(p,(x,\varphi(LJR)),q)$, we do not change the transduction, since elements
outside of $\E{M*\B}$ cannot occur in a product that results in $1$.
Therefore, we assume that every edge of $A$ is of the form $(p,(x,\varphi(LJR)),q)$ as in \eqref{eq:xdist}.

In order to be able to denote several appearing rational sets using a pair of
states, we construct finite automata
\begin{align*}
B^{(-)}&=(Q^{(-)},X',E^{(-)},q_0,\emptyset), \\
B^{(0)}&=(Q^{(0)},Y\cup Z,E^{(0)},q_0,\emptyset), \\
B^{(+)}&=(Q^{(+)},X',E^{(+)},q_0,\emptyset)
\end{align*}
such that for each edge $(p,(x,\varphi(LJR)),q)\in E$, we have
$L=L_{r,s}(B^{(-)})$, $J=L_{t,u}(B^{(0)})$, and $R=L_{v,w}(B^{(+)})$ for some
states $r,s\in Q^{(-)}$, $t,u\in Q^{(0)}$, $v,w\in Q^{(+)}$. Because of \eqref{eq:xdist}, we can assume that in these automata 
there are subsets $\tilde{Q}^{(-)}\subseteq Q^{(-)}$,
$\tilde{Q}^{(+)}\subseteq Q^{(+)}$ such that in $B^{(-)}$, an edge
is labeled $\bar{x}$ if and only if it enters a state in $\tilde{Q}^{(-)}$
and an edge in $B^{(+)}$ is labeled $x$ if and only if it leaves a state in
$\tilde{Q}^{(+)}$.  For each $r,s\in \tilde{Q}^{(-)}$, $t,u\in Q^{(0)}$, $v,w\in
\tilde{Q}^{(+)}$, let
\[ L_{r,s}=\varphi(L_{r,s}(B^{(-)})),~~~J_{t,u}=\varphi(L_{t,u}(B^{(0)})),~~~R_{v,w}=\varphi(L_{v,w}(B^{(+)})), \]
\[ \tilde{L}_{r,s}=\{\varphi(w) \mid w\in(Y\cup Z)^*,~ w\bar{x}\in L_{r,s}(B^{(-)}) \}, \]
\[ \tilde{R}_{v,w}=\{\varphi(w) \mid w\in(Y\cup Z)^*,~ xw\in L_{v,w}(B^{(+)}) \}. \]
By \eqref{eq:xdist}, every edge in $A$ is of the form $(p,(x,L_{r,s}J_{t,u}R_{v,w}),q)$.

The essential idea of the construction is to maintain a representation of a set
of possibly reached configurations. Roughly speaking, we represent a sequence
of rational subsets of $\Ri{M*\B}\times C$ by elements of $M*\B^{(n)}$. Now in
order to simulate the multiplication of a set of the form $L_{r,s}$, we have to
output a set of elements of $C$ that appear as output while canceling out
elements on the stack with those in $L_{r,s}$.  Therefore, we will output sets
of the form
\[ C_{v,w,r,s} = \{ c\in C \mid (1,c) \in R_{v,w}L_{r,s} \}. \]
By Lemma \ref{lemma:semilinear}, these sets are semilinear. 

In the course of a computation, we will have to simulate the multiplication of
rational subsets of $M\times C$. To this end, we will use the hypothesis of
$M$ being \lelim{} in the following way. Assume that 
\[ W~=~ \tilde{Q}^{(-)}\times \tilde{Q}^{(-)} ~~\cup~~Q^{(0)}\times Q^{(0)} ~~\cup~~\tilde{Q}^{(+)}\times \tilde{Q}^{(+)}\]
is an alphabet. Let $D=(\{q\},W,M,C,E',q,\{q\})$ be the rationally labeled
valence transducer over $M$ with output in $C$ with the following edges:
\begin{itemize}
\item for each $r,s\in \tilde{Q}^{(-)}$, create a loop on $q$ with input $(r,s)\in W$ and label $\tilde{L}_{r,s}$,
\item for each $t,u\in Q^{(0)}$, create a loop on $q$ with input $(t,u)\in W$ and label $J_{t,u}$, and
\item for each $v,w\in \tilde{Q}^{(+)}$, create a loop on $q$ with input $(v,w)\in W$ and label $\tilde{R}_{v,w}$.
\end{itemize}
Since $M$ is \lelim{}, we can transform $D$ into \anEmpty{} $\emptyWord$-free valence transducer $\hat{D}=(\hat{Q},X,M,\SL{C},\hat{E},q_0,\hat{F})$
over $M$ with output in $\SL{C}$ such that $\SLMap{T(\hat{D})}=T(D)$.

As mentioned above, we will encode rational subsets of $\Ri{M*\B}\times C$ by
elements of $M*\B^{(n)}$. The monoid structure of $M*\B^{(n)}$ allows us to
use the positive generators of the $n$ instances of $\B$ as stack symbols.
Specifically, for every pair of states $v,w\in\tilde{Q}^{(+)}$, we will have a symbol
$\PosR{v}{w}$ that represents the set $R_{v,w}$. 

First suppose that sets $J_{t,u}$ do not occur on edges. When an element of $R_{v,w}$ is completely
canceled out by an element of $L_{r,s}$, then we have to output an element of $C_{v,w,r,s}$.
Therefore, when $\PosR{v}{w}$ is on top and we want to simulate the multiplication of $L_{r,s}$,
we output the semilinear set $C_{v,w,r,s}$ and remove $\PosR{v}{w}$.

By construction, composing an element of $R_{v,w}$ with one of $L_{r,s}$ always
yields one whose first component is in
some $\pi_1(R_{v,w'})$, $w'\in \tilde{Q}^{(+)}$ or one outside of $\E{M*\B}\times C$.  Therefore, in order to
simulate a computation where an element of $L_{r,s}$ cancels out only part of
an element of $R_{v,w}$, we have a \emph{split} operation, which removes a
symbol $\PosR{v}{w}$ from the top and puts $\PosR{v}{w'}\PosR{w'}{w}$ in its
place, so that the simulation of $L_{r,s}$ can then cancel out $\PosR{w'}{w}$
and output $C_{w',w,r,s}$.  Note that the set $C_{w',w,r,s}$ contains only the
outputs for those compositions where the elements actually cancel out. In particular,
those compositions that yield elements outside of $\E{M*\B}\times C$ provide no
output in $C_{w',w,r,s}$.

In order to simulate an element of $L_{r,s}$ that cancels out an element in the
composition of the two topmost rational sets, we need a way to merge two representations of
rational sets. However, if we would merge two representations of rational
subsets of $\Ri{M*\B}\times C$ into one, the resulting representation would not
be of the form $\PosR{v}{w}$, since it has to keep track of what states have to
be visited on the way. Furthermore, the more representations we would merge,
the more information we would have to maintain. 

Therefore, we will not merge representations of the form $\PosR{v}{w}$.
Instead, we have another kind of symbols: the symbol $\NegR{r}{s}$ stands for
an element of $\Ri{M*\B}\times C$ that can be canceled out by one of
$L_{r,s}$.  Furthermore, the occurrence of such a symbol also implies that the
corresponding output of the canceling process has already been performed. This
means, the symbol $\NegR{r}{s}$ is produced by an operation \emph{cancel} that
removes $\PosR{v}{w}$, places $\NegR{r}{s}$ on top and outputs $C_{v,w,r,s}$.
Since $C$ is commutative, this early output does not change the result.\todo{Note that representations $\NegR{r}{s}$ can be safely merged without loss of information, etc.}  The
\emph{merge} operation then consists of removing $\NegR{r}{s}\NegR{s}{s'}$ and
putting $\NegR{r}{s'}$ in its place. Since we will always be able to assume that a symbol
$\PosR{v}{w}$ has already been turned into a $\NegR{r}{s}$, we can always
simulate the application of a set $L_{r,s}$ by removing $\NegR{r}{s}$.

Finally, we have to simulate the application of sets $J_{t,u}$. To this end, we use an
edge $(p,(t,u),m,S,q)$ in the transducer $\hat{D}$. The state information of $\hat{D}$ 
is then stored in symbols $\State{p}$ on the stack. Thus, we simulate $J_{t,u}$ by removing
$\State{p}$ from the stack, using $S$ as output, and adding $m\State{q}$ on the stack.

In order to let elements of $M$ that are factors of elements in $R_{v,w}$,
i.e., elements of $\tilde{R}_{v,w}$, interact with sets $J_{r,s}$, we have two
further operations: \emph{convert-to} and \emph{convert-from}.
Convert-to-$M$ removes an element $\PosR{v}{w}$ from the stack and instead adds
$\Between m\State{q}$ on the stack and outputs $S$, where $(q_0,(v,w),m,S,q)$
is an edge in $\hat{D}$.  That is, the element represented by $\PosR{v}{w}$ can be thought of as being
handed over to $\hat{D}$. Here, $\Between$ represents the $a$ that was part of
$R_{v,w}$, but not of $\tilde{R}_{v,w}$.  Thus, convert-from-$M$ initiates a
subsequence of stack elements that simulate a computation of $\hat{D}$. On the
other hand, convert-from-$M$ will terminate such a subsequence by simulating the multiplication of a
set of the form $\tilde{L}_{r,s}$. It removes $\State{q}$, adds $m$, removes
$\Between$, adds $\NegR{r}{s}$, and outputs $S$, where $(q,(r,s),m,S,q_f)$ is an
edge in $\hat{D}$ and $q_f$ is a final state of $\hat{D}$.

Formally, let $\Theta$ be the alphabet
\[ \Theta= \{\PosR{v}{w}, \NegR{r}{s}, \State{q}, \Between \mid v,w\in \tilde{Q}^{(+)},~ r,s\in\tilde{Q}^{(-)},~ q\in\hat{Q} \} \]
and let $n=|\Theta|$. We let each of the symbols $x\in\Theta$, together with
its counterpart $\bar{x}$, be the generators of one of the instances of $\B$ in
$M*\B^{(n)}$.  Sometimes, it is necessary to apply one of the aforementioned operations
not on top of the stack, but one one symbol below the top. Therefore, for
the operations \emph{split}, \emph{merge}, and \emph{cancel}, we have a
\emph{deep} variant, which nondeterministically removes some $x\in\Theta$, then
performs the original operation and then puts $x$ back on top.

\mathchardef\mhyphen="2D
\newcommand{\OpSplit}{\Op{split}}
\newcommand{\OpMerge}{\Op{merge}}
\newcommand{\OpConvertTo}{\Op{convert\mhyphen{}to}}
\newcommand{\OpConvertFrom}{\Op{convert\mhyphen{}from}}
\newcommand{\OpCancel}{\Op{cancel}}
\newcommand{\OpDeepSplit}{\Op{deep\mhyphen{}split}}
\newcommand{\OpDeepCancel}{\Op{deep\mhyphen{}cancel}}
\newcommand{\OpDeepMerge}{\Op{deep\mhyphen{}merge}}
\newcommand{\OpDeepConvertFrom}{\Op{deep\mhyphen{}convert\mhyphen{}from}}
\newcommand{\OpDeepDeepMerge}{\Op{deep\mhyphen{}deep\mhyphen{}merge}}
\newcommand{\OpDeepDeepCancel}{\Op{deep\mhyphen{}deep\mhyphen{}cancel}}

Formally, an operation is a (finite) set of elements of
$(X\cup\{\emptyWord\})\times (M*\B^{(n)})\times \SL{C}$.
In accordance with the explanation above, we have the following operations:
\begin{itemize}
\item $\OpSplit=\{(\emptyWord, \overline{\PosR{v}{w}}\PosR{v}{v'}\PosR{v'}{w}, \{1\}) \mid v,v',w\in\tilde{Q}^{(+)} \}$
\item $\OpDeepSplit=\{(\emptyWord, \bar{x}sx, S) \mid x\in\Theta,~(\emptyWord, s, S)\in\OpSplit \}$
\item $\OpMerge=\{(\emptyWord, \overline{\NegR{r}{r'}}\overline{\NegR{r'}{s}}\NegR{r}{s}, \{1\}) \mid r,r',s\in\tilde{Q}^{(-)} \}$
\item $\OpDeepMerge=\{(\emptyWord, \bar{x}sx, S) \mid x\in\Theta,~(\emptyWord, s, S)\in\OpMerge \}$
\item $\OpDeepDeepMerge=\{(\emptyWord, \bar{x}sx, S) \mid x\in\Theta,~(\emptyWord, s, S)\in\OpDeepMerge \}$
\item $\OpConvertTo=\{(\emptyWord, \overline{\PosR{v}{w}}\Between m \State{q},S) \mid v,w\in Q^{(+)},~(q_0,(v,w),m,S,q)\in\hat{E} \}$
\item $\OpConvertFrom=\{(\emptyWord, \overline{\State{q}}m\overline{\Between}\NegR{r}{s},S) \mid r,s\in Q^{(-)},~(q,(r,s),m,S,q_f)\in\hat{E},~q_f\in\hat{F} \}$
\item $\OpDeepConvertFrom=\{(\emptyWord, \bar{x}sx, S) \mid x\in\Theta,~(\emptyWord, s, S)\in\OpConvertFrom \}$
\item $\OpCancel=\{(\emptyWord, \overline{\PosR{v}{w}}\NegR{r}{s}, C_{v,w,r,s}) \mid v,w\in\tilde{Q}^{(+)},~r,s\in\tilde{Q}^{(-)} \}$
\item $\OpDeepCancel=\{(\emptyWord, \bar{x}sx, S) \mid x\in\Theta,~(\emptyWord, s, S)\in\OpCancel \}$
\item $\OpDeepDeepCancel=\{(\emptyWord, \bar{x}sx, S) \mid x\in\Theta,~(\emptyWord, s, S)\in\OpDeepCancel \}$
\end{itemize}

We will now describe the transducer $\hat{A}$ in detail. Although $\hat{A}$
will have $\emptyWord$-transitions, we will argue later that 
every element of $T(A)$ can be accepted by $\hat{A}$ using a
computation that uses only a bounded number of $\emptyWord$-transitions before and after every
non-$\emptyWord$-transition. Thus, it is clearly possible to transform $\hat{A}$ into an
equivalent $\emptyWord$-free valence transducer over $M*\B^{(n)}$.

$\hat{A}$ is obtained from $A$ by removing all edges and then for each edge $(p,x,L_{r,s}J_{t,u}R_{v,w},q)$, gluing in the automaton
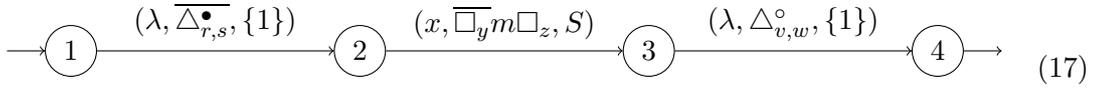
\begin{equation}\label{eq:gluedin}
\begin{tikzpicture}[every state/.style={minimum size=10pt}]
\node[state, initial, initial text=]  (a) at (0,0) {1};
\node[state]                          (b) at (3.8,0) {2};
\node[state]                          (c) at (7.6,0) {3};
\node[state,accepting by arrow]       (d) at (11.4,0) {4};
\path[->] (a) edge node [above] {$(\emptyWord, \overline{\NegR{r}{s}}, \{1\})$} (b)
          (b) edge node [above] {$(x, \overline{\State{y}}m\State{z}, S)$} (c)
          (c) edge node [above] {$(\emptyWord, \PosR{v}{w}, \{1\})$} (d);
\end{tikzpicture}
\end{equation}
between $p$ and $q$ for every edge $(y, (t,u), m, S, z)$ in $\hat{D}$.
Furthermore, on every state of $\hat{A}$ (including those in the glued in
automata), we add loops labeled with the operations defined above. 
Finally, we add a loop labeled $(\emptyWord,\NegR{r}{r},\{1\})$ for each $r\in
\tilde{Q}^{(-)}$ on the initial state and a loop labeled $(\emptyWord,
\overline{\PosR{v}{v}}, \{1\})$ for each $v\in \tilde{Q}^{(+)}$ on each final
state.

By the definition, it is clear that $\SLMap{T(\hat{A})}\subseteq T(A)$. On the other
hand, we can accept every pair $(w,c)\in T(A)$ by $\hat{A}$ in the following
way. First we bring a symbol $\NegR{r}{r}$ on the stack to represent an empty storage. 
The first simulated edge $(p, x, L_{r,s}J_{t,u}R_{v,w}, q)$, which has $s=r$, will thus be able to 
take the edge from state 1 to 2 in \ref{eq:gluedin}.
We assume that on the stack, there are no symbols of the form
$\NegR{r}{s}$ except for one representing the empty stack. Thus, the monoid element in the configuration is contained in 
\[ \{\PosR{v}{w},~\Between{}m\State{y}\mid v,w\in\tilde{Q}^{(+)},~ m\in M,~ y\in\hat{Q}\}^*~ \cup~ \{\NegR{r}{r} \mid r\in \tilde{Q}^{(-)}\}.\]
For each edge $(p, x, L_{r,s}J_{t,u}R_{v,w}, q)$ in the computation in $A$:
\begin{enumerate}
\item \label{it:phase:unbounded} Apply a sequence of $\OpCancel$/$\OpDeepCancel$, $\OpConvertFrom$/$\OpDeepConvertFrom$, $\OpMerge$, and
$\OpSplit$/$\OpDeepSplit$
loops in state $1$ to obtain the symbol $\NegR{r}{s}$ on top of the stack. Note
that for this, we need to use a $\OpSplit$ or $\OpDeepSplit$ loop at most once,
namely for the lowest used occurrence of a $\PosR{v'}{w'}$, which might be 
canceled only partially.
\item \label{it:phase:left} Use the edge $(\emptyWord,\overline{\NegR{r}{s}},\{1\})$ in \eqref{eq:gluedin}.
\item \label{it:phase:split} If necessary, use a $\OpSplit$ loop in state $2$.
\item \label{it:phase:convert} If necessary, use a $\OpConvertTo$ loop in state $2$.
\item \label{it:phase:d} Choose an edge $(x,\overline{\State{y}}m\State{z},S)$ in \eqref{eq:gluedin}.
\item \label{it:phase:right} Use the edge $(\emptyWord,\PosR{v}{w},\{1\})$ in \eqref{eq:gluedin}.
\end{enumerate}

Note that the only one of these phases which uses an unbounded number of operations is the first one. 
Therefore, we will change the computation by moving the operations of this phase to the point where the modified symbols are created. 
Thereby, we guarantee that
before the application of $(\emptyWord,\overline{\NegR{r}{s}},\{1\})$ we only need a bounded number of steps.
This is done as follows. In each of the phases \ref{it:phase:left} through \ref{it:phase:right}:
\begin{enumerate}[label=(\roman*)]
\item\label{rule:cancel} After each introduction of a $\PosR{v}{w}$ (by a $\OpSplit$/$\OpDeepSplit$ or by adding $\PosR{v}{w}$ directly): if this occurrence is eventually canceled (without being split), cancel it now.
This can be done using $\OpDeepCancel$ or $\OpDeepDeepCancel$.
\item\label{rule:convert} After each application of a $\overline{\State{y}}m\State{z}$: if the corresponding subsequence $\Between{}m'\State{z}$ is eventually converted (without adding another $\overline{\State{z}}m''\State{z'}$), convert it now.
This can be done using $\OpConvertFrom$.
\item \label{rule:merge} Whenever a symbol $\NegR{r}{s}$ produced by \ref{rule:cancel} or \ref{rule:convert} is eventually merged with a symbol below it, merge them now.
This can be done using $\OpMerge$, $\OpDeepMerge$, or $\OpDeepDeepMerge$.
\end{enumerate}
If we also obey these rules in the in the first phase, we can assume that any
$\PosR{v}{w}$ or $\Between{}m\State{z}$-subsequence that was
canceled/converted and then merged with the current underlying symbol in the old
computation, is now already merged.  Therefore, we can change the first phase so as to 
do only a bounded number of operations to obtain $\NegR{r}{s}$ on top of the stack.
Note that obeying rules \ref{rule:cancel} through \ref{rule:merge} will not yield an unbounded number of
operations, since in the first phase, there is at most one occurrence of $\OpSplit$ or $\OpDeepSplit$.

In the end, the stack should contain a symbol $\PosR{v}{v}$, $v\in\tilde{Q}^{(+)}$, to represent the empty storage. This can then
by removed by the loop labeled $(\lambda, \overline{\PosR{v}{v}}, \{1\})$ on the final state.

Thus, any $(w,c)\in T(A)$ can be produced by a computation in $\hat{A}$ using only a bounded number of
$\emptyWord$-transitions before and after any input symbol. 
Hence, $\hat{A}$ can be easily transformed into an equivalent valence transducer with
no $\emptyWord$-transitions.
\end{proofqed}

We are now ready to prove the first main result.
\begin{proofqedtitle}{Proof of Theorem \ref{theorem:firstmain}}
Clearly, \ref{item:firstmain:cs} and \ref{item:firstmain:np} each imply \ref{item:firstmain:decidable}. Lemma
\ref{lemma:undecidable} shows that \ref{item:firstmain:decidable} implies
\ref{item:firstmain:nopath}.  By Lemma \ref{lemma:graph}, \ref{lemma:b},
\ref{lemma:z}, and \ref{lemma:stack}, 
\ref{item:firstmain:nopath} implies that $\M\Gamma$ is \lelim{}. By Lemma \ref{lemma:stronger}, this implies
\ref{item:firstmain:equal}. Finally, by Lemma \ref{lemma:npcs},
\ref{item:firstmain:equal} implies \ref{item:firstmain:cs} and \ref{item:firstmain:np}.
\end{proofqedtitle}

We will now prove Theorem \ref{theorem:secondmain}.
By Theorem \ref{theorem:firstmain}, we already know that when $r\le 1$, we have
$\VA{\M\Gamma}=\VAl{\M\Gamma}$.  Hence, we only have to show that
$\VA{\M\Gamma}\subsetneq\VAl{\M\Gamma}$ if $r\ge 2$. Greibach \cite{Greibach1978} and, independently, Jantzen \cite{Jantzen1979a,Jantzen1979b}
have shown that the language
\[ L_1=\{wc^n \mid w\in \{0,1\}^*,~ n\le \bin(w) \}, \]
can be accepted by a partially blind counter machine with two counters, but not
without $\emptyWord$-transitions. Here, $\bin(w)$ denotes the number obtained by
interpreting $w$ as a base $2$ representation:
\[ \bin(w1)=2\cdot \bin(w)+1,~~~\bin(w0)=2\cdot \bin(w),~~~\bin(\emptyWord)=0. \]
Since we have to show $\VA{\B^r\times\Z^s}\subsetneq\VAl{\B^r\times\Z^s}$ and
we know $L_1\in\VAl{\B^r\times\Z^s}$, it suffices to prove
$L_1\notin\VA{\B^r\times\Z^s}$. We do this by transforming Greibach's proof into a
general property of languages accepted by valence automata without
$\emptyWord$-transitions. We will then apply this to show that
$L_1\notin\VA{\B^r\times\Z^s}$.

\begin{definition}
Let $M$ be a monoid. For $x,y\in M$, write $x\equiv y$ iff $x$ and $y$ have the
same set of right inverses.  For a finite subset $S\subseteq M$  and $n\in\N$,
let $f_{M,S}(n)$ be the number of equivalence classes of $\equiv$ in
$S^n\cap\Ri{M}$.
\end{definition}

The following notion is also used as a tool to prove lower bounds in state
complexity of finite automata \cite{GlaisterShallit1996}. Here, we use it to
prove lower bounds on the number of configurations that an automaton must be able to
reach in order to accept a language $L$.
\begin{definition}
Let $n\in\N$. An \emph{$n$-fooling set for a language $L\subseteq\Theta^*$} is a set $F\subseteq \Theta^n\times\Theta^*$ such that
\begin{itemize}
\item for each $(u,v)\in F$, we have $uv\in L$, and
\item for $(u_1,v_1), (u_2,v_2)\in F$ such that $u_1\ne u_2$, we have
$u_1v_2\notin L$ or $u_2v_1\notin L$.
\end{itemize}
The function $g_L:\N\to\N$ is defined as
\[ g_L(n)= \max\{|F| \mid \text{$F$ is an $n$-fooling set for $L$}\}. \]
\end{definition}

\begin{lemma}\label{lemma:lbound}
Let $M$ be a monoid and $L\in\VA{M}$. Then there is a constant $k\in\N$ and a
finite set $S\subseteq M$ such that $g_L(n)\le k\cdot f_{M,S}(n)$ for all
$n\in\N$.
\end{lemma}
\begin{proofqed}{lemma:lbound}
Let $k$ be the number of states in the automaton for $L$ and $S$ be the
elements appearing on edges.  Suppose $g_L(n) > k\cdot f_{M,S}(n)$ for some $n$
and let $F=\{(u_1,v_1),\ldots,(u_m,v_m)\}$ be an $n$-fooling set for $L$,
with $m>k\cdot f_{M,S}(n)$.  Since $u_iv_i\in L$ for $1\le i\le m$, we have an
accepting computation for each of these words. Let $(q_i,x_i)$ be the configuration
reached in such a computation after reading $u_i$, $1\le i\le m$. Since the automaton
has no $\emptyWord$-transitions, we have $x_i\in S^n$ for any $1\le i\le m$. Moreover, since
a final configuration is reachable from $(q_i,x_i)$, we also have $x_i\in\Ri{M}$.
Furthermore, since $m>k\cdot f_{M,S}(n)$, there are indices $i\ne j$ with $q_i=q_j$
and $x_i\equiv x_j$. This means however, that $u_iv_j\in L$ and $u_jv_i\in L$,
contradicting the fooling set condition.
\end{proofqed}

\begin{lemma}\label{lemma:foolingsets}
For $L=L_1$, we have $g_L(n)\ge 2^n$ for any $n\in\N$.
\end{lemma}
\begin{proofqed}{lemma:foolingsets}
Let $n\in\N$, and let $F$ consist of all $(u,v)$ such that $u\in\{0,1\}^n$ and
$v=c^{\bin(u)}$. Then, $F$ is an $n$-fooling set for $L$ with $|F|\ge 2^n$: we
have $uv\in L$ for any $(u,v)\in F$. Furthermore, if $u\ne u'$ for
$(u,v),(u',v')\in F$, assume $\bin(u)<\bin(u')$. Then $uv'\notin L$.
\end{proofqed}

\begin{lemma}\label{lemma:polybound}
Let $M=\B^r\times\Z^s$ for $r,s\in\N$ and $S\subseteq M$ a finite set. Then
$f_{M,S}$ is bounded by a polynomial.
\end{lemma}
\begin{proofqed}{lemma:polybound}
Every $x\in\B$ can be written uniquely as $x=\bar{a}^ka^\ell$. We define $|x|=\ell$. 
For $y\in\Z$, we have the usual absolute value $|y|$. Thus, for $z\in\B^r\times\Z^s$ and $z=(x_1,\ldots,x_r,y_1\ldots,y_s)$ we can define
\[ |z|=\max\{|x_i|, |y_j|\mid 1\le i\le r, ~1\le j\le s\}. \]
Let $m=\max\{|x| \mid x\in S\}$. Then, for $z\in S^n\cap\Ri{M}$, we have $|z|\le m\cdot n$.
Since every element in $\Ri{M}$ is of the form
$(a^{\ell_1},\ldots,a^{\ell_r},y_1,\ldots,y_s)$, we have
\[ f_{M,S}(n)\le |S^n\cap \Ri{M}|\le (m\cdot n+1)^r\cdot (2\cdot m\cdot n+1)^s, \]
which is a polynomial in $n$.
\end{proofqed}

\begin{proofqedtitle}{Proof of Theorem \ref{theorem:secondmain}}
If $r\le 1$, Theorem \ref{theorem:firstmain} already implies that $\VA{\M\Gamma}=\VAl{\M\Gamma}$.
If $r\ge 2$, we have $L_1\in\VAl{\M\Gamma}$, but Lemmas \ref{lemma:lbound}, \ref{lemma:foolingsets}, and
\ref{lemma:polybound} together imply that $L_1\notin\VA{\M\Gamma}$.
\end{proofqedtitle}

\emphasize{Acknowledgements}
The author would like to thank Nils Erik Flick, Reiner H\"{u}chting, Matthias Jantzen, and Klaus Madlener for comments that
improved the presentation of the paper.

\bibliographystyle{plain}
\bibliography{bibliography}
\end{document}